\documentclass[12pt,preprint]{aastex}

\shorttitle{A multi-zone  model for SSC blazar variability}
\shortauthors{Graff, Georganopoulos, Perlman, Kazanas}

\begin{document}

\title{A multi-zone  model  for simulating the high energy variability of TeV blazars}

\author{Philip B. Graff \altaffilmark{1}, Markos Georganopoulos \altaffilmark{1,2}, Eric S. Perlman \altaffilmark{3}, Demosthenes Kazanas \altaffilmark{2}}

\altaffiltext{1}{Department of Physics, Joint Center for Astrophysics, University of Maryland, Baltimore County, 1000 Hilltop Circle, Baltimore, MD 21250, USA}
\altaffiltext{2}{NASA Goddard Space Flight Center, Code 663, Greenbelt, MD 20771, USA}
\altaffiltext{3}{Department of Physics and Space Sciences, Florida Institute of Technology, 150 West University Boulevard, Melbourne, FL 32901, USA}

\begin{abstract}

 We present a time-dependent  multi-zone code for simulating the variability of Synchrotron-Self Compton (SSC) sources. The code adopts a  multi-zone pipe geometry for the emission region,
 appropriate for simulating emission from a standing or propagating shock in a collimated jet.
 Variations in the injection of relativistic electrons in the inlet propagate along the length of  the pipe cooling radiatively. Our  code for the first time takes into account the non-local, time-retarded nature of synchrotron self-Compton (SSC) losses that are thought to be dominant in  TeV blazars.
  The observed synchrotron and SSC emission is followed self-consistently  taking into account light travel time delays. At any given time, the emitting  portion of the pipe depends on the frequency and the nature of the  variation followed.
Our simulation 
employs only one additional physical parameter relative to one-zone models, that of the pipe length and
 is computationally very efficient, using  simplified expressions for the SSC processes. The code 
 will be useful for observers modeling GLAST,  TeV, and X-ray observations of SSC blazars.

\end{abstract}


\section{Introduction \label{section:intro}}

In blazars, radio loud active galaxies with their relativistic jets pointing close to our  line of sight  \citep{blandford78}, the observed radiation is dominated by  relativistically beamed emission from the sub-pc base  of the jet. The blazar spectral energy distribution (SED)  consists of two components. The first one, peaking at sub-mm to X-ray energies is almost certainly due to synchrotron radiation, while the second one peaking at MeV to TeV energies is believed to be of inverse Compton (IC) nature, with both components produced by  the same population of relativistic electrons. The nature of the IC-scattered seed photons   is still not clear, with both external optical-UV photons from the broad line region \citep{sikora94} and IR photons from the putative molecular torus \citep{blazejowski00}, as well as  synchrotron photons (SSC, e.g. Maraschi, Ghisellini \& Celotti 1992) contributing. It is believed
that  in the case of powerful blazars peaking at MeV to GeV energies, external seed photons  from the broad line region dominate  the IC scattering, while for weaker lineless  blazars, peaking at $\sim$ TeV energies,
SSC is the dominant emission mechanism. Recent observational results (e.g. D'Arcangelo et al. 2007; Marscher et al. 2008), however, place the blazar 
emission site beyond the broad line region, lending support  to the possibility that even in powerful blazars the GeV emission process  may be pure SSC.
For a review of  leptonic models, as well as hadronic models for blazar emission (e.g.  Aharonian 2000)   see \citet{boettcher06}.

Due to the small angular size of the blazar emission region,  it is not possible to spatially resolve the emitting region. Because of this, information about the structure of the emitting source can be obtained
only through  multiwavelength variability studies. Particularly telling is the variability of the emission produced by  the highest energy electrons, because these electrons lose energy very quickly and  exist only close to the sites where they have been produced. 
The goal of multiwavelength variability  campaigns, involving in many cases observations from radio up to TeV energies,  is to study the  characteristics of blazar variability, such as correlations and/or  time delays between  different energies, spectral characteristics of the observed variability,  and the amplitude of  variability as a function of energy.

 Most notable amongst blazars are the so called TeV blazars  for which the synchrotron emission peaks at X-ray energies and the SSC emission peaks at TeV energies, as they present the active galaxies  producing the highest confirmed electron energies.
 Variations of TeV blazars in these two bands can be extremely  rapid (TeV doubling times as short as a few min; Aharonian et al. 2007), suggesting highly relativistic sub-pc scale  flows (Doppler factors $\delta\sim 50$;  e.g. Begelman, Fabian \& Rees 2008) that decelerate substantially  \citep{georganopoulos03,ghisellini05} to match the much slower speeds required by VLBI observations \citep{piner04, piner08}.  The TeV and X-ray variations are usually well correlated (e.g. Fossati et al. 2008; Maraschi et al. 1999; Sambruna et al. 2000),  as expected, because they present variations by the same electron population.
 Usually, the lower energy emission  within each of these bands peaks with small time delays relative to the higher energy emission (e.g. Fossati et al. 2000), while  the X-ray and TeV spectra become harder with increasing flux (e.g.  Takahashi et al. 1996).
 In certain  cases, however, the X-ray and TeV variability do not seem to be correlated in a simple way
 (e.g. Aharonian et al. 2005). An intriguing  variability pattern is that of the so-called `orphan' flares,  rare TeV flares  that are not accompanied by X-ray  flares (e.g. Krawczynski et al. 2004,  B\l a\.zejowski et. al. 2005). While the correlated X-ray - TeV flares can be understood through an increase of the high energy emitting electrons, orphan TeV flares defy such a straightforward explanation.

Models of  blazar emission to date have, for the most part,  been in some form of homogeneous one zone models (e.g. Mastichiadis \& Kirk 1997; Krawczynski,  Coppi, \&  Aharonian  2002).  Such models, although  appropriate for modeling the steady-state emission of a source, cannot simulate variability faster than the zone light crossing time.
The basic limitation of one zone models stems from the fact that  the high energy variability of both the synchrotron and SSC components is produced by high energy electrons with cooling times shorter than the light crossing time. Even if  we assume that a disturbance in the radiating plasma (e.g. a higher density)  instantaneously propagates
across the zone,  the received radiation would be smeared out for timescales shorter than the light crossing time, due to light travel time delays from different parts of the source (\S 
\ref{section:testing}; also Chiaberge \& Ghisellini 1999), and no variability faster than the light crossing time would be observed.
One, therefore, cannot use one zone models to infer the source structure from high energy variability.  

Inhomogeneous variability models of increasing degree of sophistication have attempted to overcome
the problems of one-zone models. The basic idea is to overcome the unphysical instantaneous injection
throughout the source by adopting a specific geometry for the plasma flow that includes an inlet for injecting the radiating plasma. Variations in the injected plasma  propagate and produce variations in the emissivity. Calculations of  the received emission that take into account the light-travel times that radiation from
different parts of the source  takes  to reach the observer, produce light curves that, at least,  do not violate causality. How physically realistic these light curves are depends on the approximations used and on the  characteristics of the source to be modeled. For example,  synchrotron and IC losses from photons external to the source are local processes in the sense that, at a given point in the flow, the energy loss rate only depends on the local magnetic field and external photon field 
energy density, and not on the photon production throughout the  source.  This is not the case with SSC losses, because synchrotron photons produced throughout the source at earlier  times - to take into account the light travel time from one point of the source to another - contribute to the photon energy density responsible for the SSC losses and to the emissivity at a given point and time in the source. To properly model sources like TeV blazars, in which SSC losses are important or even dominant, these considerations have to be taken into account.

There have been a few  attempts during the last fifteen years to take these spatial considerations into account.  \cite{gomez94} considered a conical jet  with a  constant bulk Lorentz factor flow in which the electron plasma and the magnetic field undergo adiabatic evolution only  and 
calculated the  radio variability induced by a shock wave propagating along the jet.
\cite{georganopoulos98a,georganopoulos98b} studied a parabolic jet that hydrodynamically accelerates and focuses  to a conical geometry, and  by following the synchrotron energy losses of the emitting electrons  reproduced the  radio to X-ray  light curves  of the X-ray bright blazar PKS 2155-304. This resulted in a frequency-dependent source size, in agreement with the fact that the variability timescale of synchrotron radiation increases with decreasing frequency. It also reproduced the usually observed soft lags (variations at soft X-rays being preceded by variations at hard X-rays) and the counterclockwise X-ray flux - X-ray spectral index loops (e.g. Takahashi 1996, Maraschi 1999, Kataoka 2000, Ravasio 2004), both manifestations of radiative cooling dominating the energetics of the high energy  electrons.

\cite{kirk98} developed a semi-analytical model  in which low energy electrons are injected in a zone where they undergo acceleration and eventually escape. The acceleration zone is assumed to move with a certain velocity, leaving behind the freshly accelerated electrons that cool through synchrotron radiation.  Variations in the injection rate of low energy electrons in the acceleration zone result
in variations of the emissivity, which are integrated over the volume of the source, taking into account time delays, to produce the observed multifrequency synchrotron light curves. This model includes a treatment of particle acceleration and it is able to reproduce the uncommon hard lags (variations at hard X-rays  preceded by variations at soft X-rays) and clockwise  X-ray flux - X-ray spectral index loops \citep{zhang02,ravasio04}, both manifestations of electrons still accelerating, just before  reaching the maximum electron Lorentz factor, where the acceleration and radiative loss timescales are comparable.
It does not include, however, SSC considerations.

\cite{chiaberge99} studied  the synchrotron and SSC emission,  from a homogeneous one zone model in which they assumed an instantaneous plasma injection, but taking into account the
time delays with which the external observer would observe the variability (a similar approach was also taken by Kataoka et al. 2000). They also studied a case similar to that of Kirk et al. (1998), but without treating particle acceleration, by splitting the source into smaller one zone models that evolved autonomously, in the sense that ({\sl i}) the SSC emission inside any one of their
single zones uses as seed photons  only the synchrotron photons produced in that zone
and ({\sl ii}) the SSC energy losses in every zone are caused only by the synchrotron photons produced in that zone.  This simplified approach is a good approximation for following the energetics
of the electrons if the source is synchrotron dominated (because the SSC losses, although inappropriately calculated, are negligible), but does not produce realistic SSC light curves, because it does not calculate the emission due to upscattering synchrotron photons produced in other parts of the source
in retarded times.

A significant improvement was introduced by \cite{sokolov04} who incorporated in the calculation
of the SSC emission from a given location in an inhomogeneous source the synchrotron photons produced throughout the source in retarded times. This produces accurate SSC light curves, provided that the SSC losses  that were still treated as a local process are negligible. In a follow up paper, \cite{sokolov05},  considered also external Compton photons from the broad line region and the molecular torus. The challenge for inhomogeneous multi zone models  for sources such as  the TeV emitting blazars is  the calculation of the non-local,  time-retarded SSC losses induced by photons produced in other parts of the source.  

Here,  we present such an inhomogeneous model that, for the first time, takes into account the non-local, time-delayed  source emission on  the SSC losses. 
We assume that a power law of relativistic electrons is injected at the inlet  of a pipe, and that the electrons flow downstream and cool radiatively. Variations in the injected electron distribution propagate downstream and manifest themselves as frequency dependent variability.
This allows us to model high energy multiwavelength variability in a self-consistent manner. In \S \ref{section:onezone} we describe the one zone model, which we use  as a building block for the multizone model, and we show that, by construction, one zone models  cannot simulate variability produced by high energy electrons
with radiative cooling time shorter than the electron light crossing time from the single zone. In \S 
\ref{section:multizone} we describe our multizone, pipe-geometry model with emphasis on the 
coupling between subsequent zones and on the  calculation of the local photon field due to non-local, time-delayed emission throughout the source. This is followed in \S \ref{section:results} by a comparison of the code with analytical results and  a series of case studies. We conclude  in \S \ref{section:discussion} with a discussion of additional considerations that can be used as starting points
for future work.

\section{The One Zone Model \label{section:onezone}}

We consider a homogeneous spherical source of radius $R$ permeated by a magnetic field $B$ of energy density $B^{2}/(8\pi)$.  Energetic electrons are injected into the region at a rate 
$q(\gamma,t)$, where $\gamma$ is the electron Lorentz factor and $t$ the injection time.
These electrons lose energy through synchrotron and inverse Compton radiation and eventually escape after a characteristic time, $t_{esc}$, of the order of the light crossing time.
%This time is related to the bulk velocity   $u=R/t_{esc}$ of the electron flow in  \S \ref{section:multizone}.  
The implementation we describe is applicable to sources that are optically thin both to synchrotron emission in the frequency range under consideration and  to $\gamma$-ray absorption due to  pair-production.

The kinetic equation that describes the time-evolution of the electron energy distribution (EED) $ n(\gamma,t) $ is 
\begin{equation}
{\partial n(\gamma,t) \over \partial t}+{\partial \over \partial \gamma} [{\dot \gamma} \,n(\gamma,t)]+
{n(\gamma,t) \over t_{esc}} = q(\gamma,t).
\label{eq:kinetic}
\end{equation}
Here,  $\dot \gamma$ includes both the synchrotron losses $\dot \gamma_s$  and the inverse Compton losses $\dot \gamma_{IC}$  in the Thomson regime $(\epsilon \gamma \leq 3/4)$

\begin{equation}
\dot\gamma = \dot \gamma_{s}+\dot \gamma_{IC}, \;\;    \dot \gamma_{s}=    {4 \sigma_\tau\over 3 m c} \gamma^2 U_B, \;\;  \dot \gamma_{IC}=  {4 \sigma_\tau\over 3 m c} \gamma^2 \int_{\epsilon_{min}}^{min[\epsilon_{max},\, 3/(4\gamma)]} U(\epsilon,t)d\epsilon
\end{equation}

where $U(\epsilon,t)$ is the photon field energy  density,  $\epsilon$ is the photon energy in units of the electron rest energy $m_e c^2$, and  $\sigma_\tau$ is  the Thomson cross section.
The photon field $U(\epsilon,t)$ includes not only the synchrotron produced photons, but all the photons produced
in the source through IC scattering, including, therefore, all the higher order SSC emission.

We calculate the synchrotron emission following \cite{melrose80}:
\begin{equation}
L_s(\epsilon_s,t)=1.85 {\sqrt{2} q^3 B \over h} \int_{\gamma_{min}}^{\gamma_{max}}z^{1/3}e^{-z}n(\gamma,t)d\gamma, \;\; z=\left({2\over 3}\right)^{1/2}{\epsilon_s /B_{\star}\gamma^2},
\label{eq:melrose}
\end{equation}
where  $q$ is the electron charge, $B_\star= B / B_{crit}$, and $B_{crit} = (m_e^2 c^3) /(e \hbar)$ is the  critical magnetic field, where the electron cyclotron energy  equals its rest mass and strong field considerations become important (e.g. Harding \& Lai 2006). We note, that, although a $\delta$-function approach for calculating the synchrotron emissivity would be faster, it would 
misinterpret the spectra in cases of hard power - law injection   
$q \propto \gamma^{-p}, \; p>2 $, whose cooling is known to produce a pile-up at its high energy cutoff
 (e.g. Kardashev 1962). Also, a 
$\delta$-function synchrotron emissivity would not  produce the $f_\nu\propto \nu^{1/3}$
spectrum at frequencies below the critical frequency of the lowest energy electrons. These lower energy photons can be  important seed photons for producing hard  SSC TeV emission as \cite{katarzynski06} point out.

To obtain the SSC emission through a simple integration as in the synchrotron case, we employ the $\delta$-function approximation 
in which 
seed photons of energy $\epsilon_0$ are IC scattered by electrons of Lorentz factor $\gamma$ to  energy $\epsilon_{IC} = (4 / 3) \epsilon_0 \gamma^2$, as long as the scattering takes place in the Thomson regime ($\gamma \epsilon_0< 3/4$). If we consider seed photons in the energy range
$d\epsilon_0$ being IC scattered by electrons with Lorentz factors in the range $d\gamma$, then the
emitted IC power is $\dot \gamma_{IC} m_e c^2 n(\gamma,t) d\gamma$  and it is spread over a final photon energy
range $d\epsilon_{IC}=4(\epsilon_0\epsilon_{IC}/3)^{1/2} d\gamma$, resulting in an IC specific luminosity per seed photon energy interval
\begin{equation}
dL_{IC}(\epsilon_{IC},t)=m_ec^2 n(\gamma,t)\dot\gamma_{IC}{d\gamma \over d\epsilon_{IC}}\; \delta(\gamma-(3\epsilon_{IC}/4\epsilon_0)^{1/2}) \Theta(3/4-\gamma\epsilon_0) ,
\end{equation}
where in this context $\dot \gamma_{IC}=  (4 \sigma_\tau/ 3 m c) \gamma^2  U(\epsilon_0,t)d\epsilon_0$
and $\Theta(x)$ is the Heaviside step function.
This is written as
\begin{equation}
dL_{IC}(\epsilon_{IC},t)=  {3^{1/2}\sigma_\tau c n(\gamma,t)  U(\epsilon_0,t) d\epsilon_0 \epsilon_{IC}^{1/2}\over 4 \epsilon_0^{3/2} }  \;   \delta(\gamma-(3\epsilon_{IC}/4\epsilon_0)^{1/2}) \Theta(3/4-\gamma\epsilon_0).
\end{equation}
% In this approximation, the calculation of IC emission requires a  single integration only.
Integrating over the available seed photon distribution we obtain
\begin{equation}
L_{IC}(\epsilon_{IC},t)=  {3^{1/2}\sigma_\tau c   \epsilon_{IC}^{1/2}\over 4}   \int_{\epsilon_{0,min}}^{\epsilon_{0,max}}  n(\gamma,t)  U(\epsilon_0,t)  \epsilon_0^{-3/2}\delta(\gamma-(3\epsilon_{IC}/4\epsilon_0)^{1/2})   d\epsilon_0.
\end{equation}
The range of final photon energies $\epsilon_{IC}$ is
 $(4/3) \epsilon_{seed,min} \gamma_{min}^2<\epsilon_{IC} < \gamma_{max}$. For $\epsilon_{IC}$ within this range the limits of the above integration are:
 \begin{equation}
\epsilon_{0,min}=\displaystyle \left\{ \begin{array}{cc}
\epsilon_{seed,min}& \mbox{for $ \displaystyle \epsilon_{IC}\leq {4 \over 3}\; \epsilon_{seed,min} \gamma_{max}^2 $}\\
\\
\displaystyle {3 \epsilon_{IC} \over 4\gamma_{max}^2 } &  \mbox{for $ \displaystyle \epsilon_{IC}\geq {4 \over 3}\; \epsilon_{seed,min} \gamma_{max}^2 $}
\end{array}      
\right.,\label{eq:emin}
\end{equation}
\begin{equation}
\epsilon_{0,max}=\displaystyle \left\{ \begin{array}{cc}
\displaystyle  {3 \epsilon_{IC} \over 4\gamma_{min}^2 }&  \mbox{for $ \displaystyle \epsilon_{IC}\leq  \gamma_{min}$}\\
\\
\displaystyle { 3 \over 4 \epsilon_{IC}} &  \mbox{for $ \displaystyle \epsilon_{IC}\geq \gamma_{min} $}
\end{array}      
\right.. \label{eq:emax}
\end{equation}

Following \cite{chang70} and \cite{chiaberge99}, we discretise  the kinetic equation (\ref{eq:kinetic}), using 
a grid of logarithmically spaced Lorentz factors, $\gamma_j, \; j=0,1,2,...,j_{max}$, 
and linearly spaced time indices, $t_i$. The difference equation that describes this system is 
\begin{equation}
{n_{j,i+1}-n_{j,i} \over \Delta t} = -{\dot\gamma_{j+1,i+1}n_{j+1,i+1}- \dot\gamma_{j,i+1}n_{j,i+1} 
\over \Delta\gamma}+q_{j,i+1}-{n_{j,i+1} \over t_{esc}}.
\end{equation}
Note that this is an implicit scheme, in the sense that the calculation of  $n_{j,i+1}$ requires knowledge  not of only the previous timestep EED, but also $n_{j+1,i+1}$, the next higher $\gamma$ grid point  at the current time. It is due to the implicit nature of the numerical procedure that
this scheme is stable for large timesteps.
The difference equation  can be written as a system of tridiagonal equations
\begin{equation}
n_{j,i+1}=a \, n_{j,i}-b\, n_{j+1,i+1}+c\, q_{j,i+1},
\end{equation}
\begin{equation}
a={\Delta\gamma \over 
\Delta\gamma+\Delta t\, \Delta\gamma \, /t_{esc}-\Delta t\, \dot\gamma_{j,i+1}},\;\;b=a {\Delta t \over \Delta \gamma}\dot\gamma_{j+1,i+1},\;\; c=a\Delta t.
\end{equation}
This can be easily computed if we use the initial condition  $n_{j,0}=0 \; \forall\,  j$  (start with no relativistic electrons in the system) and the boundary condition $n_{j_{max},t }=0 \; \forall  \, t$. The first condition implies that the  initial photon field is also  zero for all photon energies.
The second condition is  satisfied if we set  $\gamma_{j_{max}} > \gamma_{max}$, because  the electrons can only lose energy, and there is no way to move to higher energies,
populating the $j_{max}$ bin of the $\gamma$-grid.
The simulation proceeds in the following manner: given $n(\gamma,t_i)$ and $U(\epsilon_0,t_i)$ we first calculate $n(\gamma,t_{i+1})$. We then calculate the synchrotron luminosity, $L_S(\epsilon,t_{i+1})$, and the inverse Compton luminosity, $L_{IC}(\epsilon,t_{i+1})$.
The specific photon energy  density $U(\epsilon_0,t_{i+1})$ for the next timestep is obtained by adding these two luminosities and dividing by $4 \pi R^2 m_e c^3$.

\subsection{Problems of one zone models in reproducing high energy variability \label{section:testing}}

By construction, in the one zone model variations in the injection propagate instantaneously throughout the source,  because no spatial coordinate enters the description of the system.
If a power - law EED,  
$q\propto \gamma^{-p}, \; \gamma_{min} \leq \gamma\leq \gamma_{max}$, is injected in the source, 
radiative cooling and electron escape  will result in a broken power law
 EED $n(\gamma)$ in the source, steepening from  an electron index $p$ to $p+1$, above $\gamma_b$,
 the electron Lorentz factor for which the escape time equals the radiative loss time
 \begin{equation}
 {\gamma_b\over \dot \gamma}=t_{esc} \Rightarrow \gamma_b={ 3 m_e c \over 4 \sigma_\tau U t_{esc}},
 \label{eq:gammab}
 \end{equation}
where $U$ stands for the total photon and magnetic field energy density in the source. It is these electrons, with $\gamma > \gamma_b$ that produce the high energy synchrotron and IC emission.
 Because the electron escape time is  of the order of the light crossing time, $t_{esc}=k t_{lc}=k R/c$, 
 $k\sim 1-$few,
 electrons with Lorentz factor $\gamma > \gamma_{lc}=k\gamma_b$, have a cooling time
  shorter than the light crossing time. 
   One therefore anticipates that even for an  injection
  event lasting much less that $t_{lc}$, the high energy variable emission produced by electrons with $\gamma>\gamma_{lc}$ will be  smeared out by light travel time effects and would appear to last for $\sim t_{lc} $,  even though in each point in the source it lasts a shorter time $\sim t_{lc} \gamma_{lc} /\gamma$.

\begin{figure}
\epsscale{0.7}
\plotone{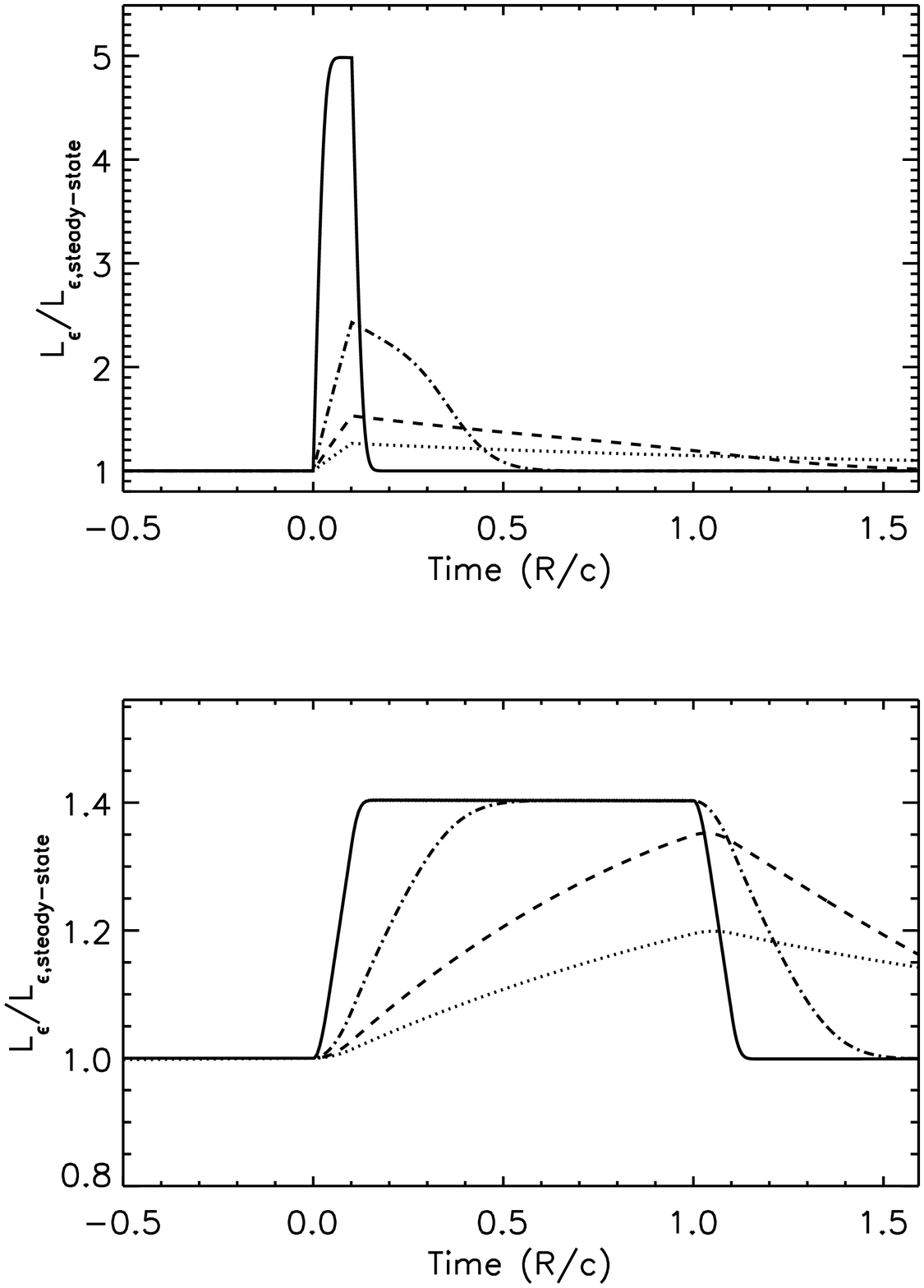}
%\plotone{f1.eps}
\caption{ A short variation by a factor of $5$ for  $t_{inj}=t_{lc}/10, \; t_{lc}=R/c$, is introduced in the one zone model. The
dotted  line tracks the synchrotron emission of the low energy electrons with cooling time $t_{cool}>t_{esc}$, the broken line of electrons with $ t_{lc} < t_{cool} <t_{esc}$, the dash-dot line of electrons with  $t_{inj} <t_{cool} < t_{lc}$, and the solid line of electrons with $t_{cool} < t_{inj} $.
In the upper frame, no light crossing delays are taken into account and this results to the unphysical result of  variability times
shorter than $t_{lc}$, for radiation produced by electrons with $t_{cool}<t_{lc}$. 
 The lower frame depicts the same event with light crossing delays taken into account. In this case, the light crossing time is the smallest observable variability scale.  The time-integrated  emission in the disturbance is the same in both cases, as can be easily seen for the highest frequency that reaches a plateau in both panels: $(5-1)\times0.1=(1.4-1)\times 1$. }
\label{fig:onezone}
\end{figure}

To demonstrate this, a flaring state was simulated by an increase by a factor of $5$ in the injection  $q(\gamma,t)$ that lasted  $t_{inj}=t_{lc}/10$.  The system was allowed to reach a steady state before the disturbance in the injection was introduced.  The emitted luminosity as a function of frequency was followed in  time allowing us to  produce light curves.  By not taking into account time delays, and wrongly assuming that at any given time  the observer  sees the entire source  as being at a single  physical state, 
  electrons with $t_{cool} < t_{lc}$   produce high energy synchrotron and SSC
  variations that last less than $t_{lc}$ (upper panel of  Figure
   \ref{fig:onezone}).

To take time delays into account, one has to  consider that if at a certain time $t$ the observer receives photons from the nearest (front)  part of the source,   slices further away from the observer will be seen as they were in retarded times  $t-r/c$,  where $r$ is the distance of the slice from the front part of the source. To treat this, at each time the luminosity of the system was recorded for a number of time steps covering a time equal to the light crossing time of the region. The luminosity observed at any time is thus the sum of the luminosity from each of these  time steps, as each one represents the light emitted by a slice sequentially further back from the observer. The resulting light
curves, plotted in the lower panel of Figure
   \ref{fig:onezone}, show that when the size of the region, and thus the time taken by light to travel across it, is accounted for, the  observed  variability of high energy electrons with $t_{cool} < t_{lc}$ is spread out over the length of the light crossing time. 
 This is similar to what \cite{chiaberge99}  observed when they performed a similar test.

%As we discuss below, by adopting a multi zone model we can reduce the size of the zones taken to be homogeneous and still model the same emitting region. This gives us the ability to accurately model high energy  variability by reducing the minimum observable variability.

 Another serious problem stemming from the lack of spatial considerations in the one-zone
 model comes from the fact that the model by construction assumes that the photons
produced in the source at a given time are instantaneously available as seed photons for IC scattering
throughout the source. This unphysical assumption has serious implications on the calculation of the SSC emissivity and on the calculation of the SSC losses, which in turn affect the evolution of the EED in the source, and through this the entire spectrum and light curves.
The first effect has been addressed by the inhomogeneous model
 of Sokolov et al. (2004) and Sokolov \& Marscher (2005). We present now the first multi-zone simulation that  incorporates the issue of light travel time effects on the SSC losses.

\section{The Multi Zone Model \label{section:multizone}}

\subsection{The  flow geometry \label{section:pipe}}
The simplest and least computationally intensive 
deviation from a homogeneous model that can address the issues discussed above is
one in which plasma is injected into a pipe of radius $R$ and length $L$ and cools radiatively as it flows downstream before it escapes after traversing  the pipe length.
Physically, this resembles the situation of a standing or propagating shock, as seen at the frame of the 
shock front.
The plasma flow velocity $u$ and the magnetic field $B$ are  constant along the pipe, and the EED is assumed  to have no lateral gradients along the cross section of the pipe. 
Relativistic plasma is injected at the base
of the flow. The injection variation timescale  in this geometry can be  arbitrarily smaller than $R/c$ without violating causality,  because, in principle, a disturbance in the plasma flow can reach the entire cross-section of the inlet at a single instance.
However,  the discretization procedure we describe below limits the range of meaningful injection variability to  timescales  greater than the plasma flow time through a zone of the flow.
 Our goal is to calculate the EED in the frame of the pipe as a function of time and distance $z$ from the inlet of the pipe, and through this  calculate the  emission received by an observer located at an angle $\theta$ to the axis of the pipe.

\subsection{The discretization of the pipe \label{section:discretization}}

 The pipe is broken down lengthwise and 
 all cells are of length $l$, comparable to the pipe radius $R$. The length of the pipe  $L=Nl$, where $N$ is the number of cells.
 Each cell is then simulated by a  one zone model. The electron injection at the first cell is  $q_1(\gamma,t)$, similar to that defined for the one zone model.  
 In each time step, the electrons that are calculated to leave each cell, are  injected  into the next cell in line in the next timestep.   The injection of electrons, therefore, in  cell $i$ at time $t_j$ is
 $q_i(\gamma,t_j)=n_{i-1}(\gamma,t_{j-1})/t_{esc}$
 and the kinetic equation for the $i$-th cell is
 \begin{equation}
{\partial n_i(\gamma,t_j) \over \partial t}+{\partial \over \partial \gamma} [{\dot \gamma} \,n_i(\gamma,t_j)]+
{n_i(\gamma,t_j) \over t_{esc}} = {n_{i-1}(\gamma,t_{j-1}) \over t_{esc}}
\label{eq:kinetic_cells}
\end{equation}
   Because in a time  $\sim t_{esc}$ the electron content of a cell is transferred to the next cell,  $t_{esc}$ is connected to the bulk flow
 velocity $u$ through $t_{esc}=l/u$. We also use  $t_{esc}$ as  the  time step of our simulation. This ensures that the actual distance a disturbance in the electron distribution  travels in a time step is equal to the bulk velocity times the time-step size, by transferring in a time step  the electron content of cell $i$ to cell $i+1$. 
 The shortest variability timescale that can be simulated by this configuration is the single cell escape time. 
 This limits  the highest energy electrons that can be followed accurately  to that of Lorentz factor
 $\gamma_b= 3 m_e cu / 4 \sigma_\tau U l$, where $U$ is the photon plus magnetic field energy density in the first cell  (see equation \ref{eq:gammab}),  and through
 this the highest energies of synchrotron and IC variations that can be reproduced.
 The advantage of the pipe configuration relative to a homogeneous model of the same size 
 is that the highest energy electron variability we can follow is not connected to the length of the entire pipe, but to the length of a single zone, a quantity that  is $N$ times shorter. This results in the pipe being able to track variations faster by  $N$, following  electrons more energetic by  $N$, and synchrotron and SSC fequencies higher by $N^2$,  relative to a homogeneous model of  size $L$. An early version of this approach was presented by Graff et al. (2007).

\subsection{The photon energy density  \label{section:photons}}
To solve the kinetic equation for each cell, an expression for the photon energy density resulting from
all other cells by taking into account light travel time delays is required. 
In general, for a region $S$ characterized by a time-dependent emission coefficient $j({\bf r'},t',\epsilon)$, the photon energy density $U({\bf r},t,\epsilon)$ is calculated by integrating   $j({\bf r'},t',\epsilon)/c$ in retarded times over the volume of the region $S$. 
Setting ${\bf r}=0$, for a point of interest in $S$, yields, without loss of generality, 
 \begin{equation}
U({\bf r}=0,t,\epsilon)={1 \over c} \int_{0}^{r'(\Omega)} j({\bf r'},t'=t-r'/c,\epsilon)dr'd\Omega.
\end{equation}
For our geometry we express this through the following approximation. Consider a  cell $i$ centered at $z_i$ being
illuminated by  a cell $m$  centered at $z_m$.  The solid angle subtended at $z_i$ by the cell $m$ is 
$\Delta\Omega\approx \pi R^2/(z_i-z_m)^2$. The photon energy density $\Delta U(z_i,z_m,t,\epsilon)$ 
at ($z_i$, t) due to photons produced at $z_j$ at retarded time $t'=t-|z_i-z_m|/c$ is:
\begin{equation}
\Delta U(z_i,z_m,t,\epsilon)={j(z_m,t'=t-|z_i-z_m|/c, \epsilon )\over c} \; {\pi R^2\over (z_i-z_m)^2} l.
\end{equation}
Making use of the fact that the volume of each cell is  $V_{c}=\pi R^2 l$, and that the luminosity $L(\epsilon)$ emitted from a cell is $L(\epsilon)=4\pi j(\epsilon) V_c$, we obtain 
\begin{equation}
\Delta U(z_i,z_m,t,\epsilon)={L(z_m,t'=t-|z_i-z_m|/c,\epsilon )\over 4\pi c  (z_i-z_m)^2}. 
\end{equation}
A summation over all cells in the pipe results to  the total photon energy density  $U(z_i,t_j,\epsilon_k)$ at cell $i$, time $t_j=jt_{esc}$, and energy $\epsilon_k$
\begin{equation}
U(z_i,t_j,\epsilon_k)={L(z_i,t_j,\epsilon_k) \over 2\pi c R(R+l)}+\displaystyle \sum_{m=0, m\ne i} ^N
{L(z_m,t'=j t_{esc}-|i-m|l/c, \epsilon_k) \over 4\pi c l^2(i-m)^2},
\end{equation}
where the first term is the photon energy density due to cell $i$ itself. 
Note that a calculation of the photon energy density requires keeping record of a data-cube of  the SED  emitted by each cell at each time for a number of time steps equal to  the light crossing time of the entire region.

\subsection{Pipe orientation \label{subsection:orientation}}

When calculating the observed luminosities, we must take into account the different distances that light must travel from each of the different cells in the pipe to the observer. This difference is
a function of the angle $\theta$ formed between the pipe and the observer. 
If photons emitted from the first cell  are received by the observer at a given time, 
photons from cell $i$ that are received simultaneously were emitted earlier by $ i\,l \cos \theta/c$.
Beaming can be easily included in this model, by assuming that the entire pipe is moving with a relativistic velocity along its axis. Then for a choice of bulk Lorentz factor $\Gamma$ and orientation angle $\theta_{obs}$  in the observer's frame, one can calculate the Doppler factor $\delta$ and the
angle $\theta$ in the frame of the pipe, and transform the arrival times by dividing by $\delta$, the observed frequencies by multiplying by $\delta$, and the observed fluxes by multiplying by $\delta^3$.

\section{Results \label{section:results}}

\subsection{Comparison with analytical results for the synchrotron dominated case.}
\begin{figure}
\epsscale{0.7}
\plotone{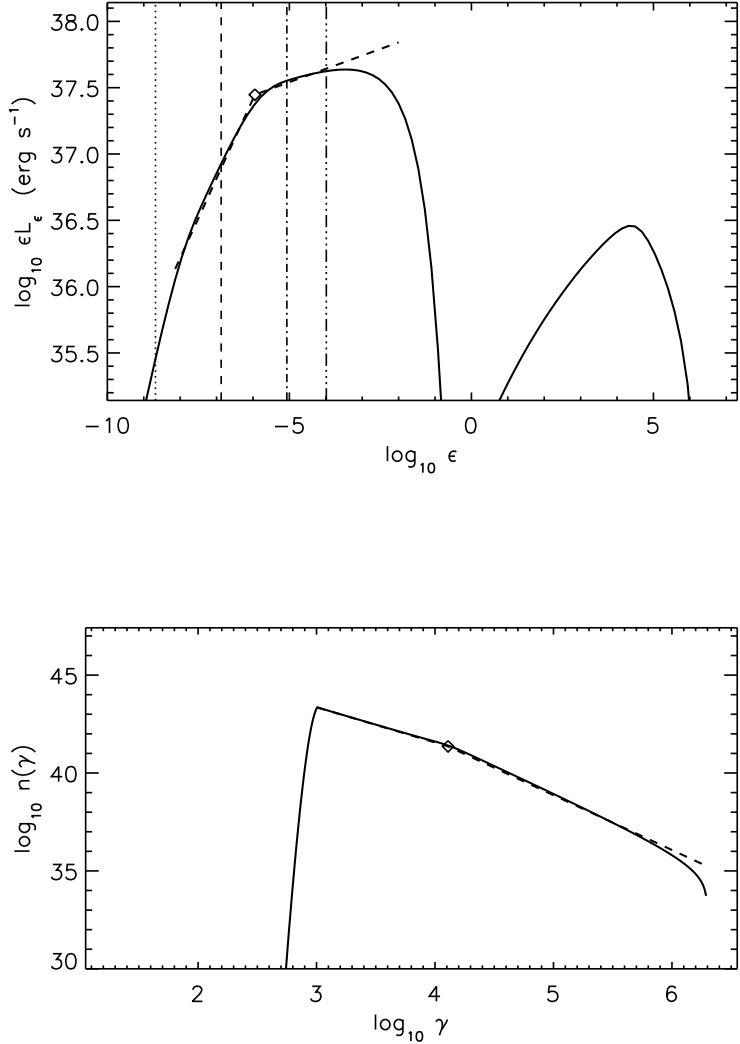}
\caption{Comparison of the steady-steady numerical result with the analytical solution of  a synchrotron dominated configuration. The following parameters have been used: A pipe of length $L=10^{16}$ cm
and radius $R=5 \times 10^{14}$ cm (aspect ratio $10:1$). The magnetic field in the pipe is $B=0.3$ G, and a power law EED is continuously injected with $\gamma_{min}=10^3$, $\gamma_{max}=2\times 10^6$, $p=1.8$. The pipe is split into 40 cylindrical slices of equal height $l=L/40=2.5\times 10^{14}$cm, and the escape time is set to $k=2$ times  the light crossing time. With these parameters fixed, the ratio of the SSC to synchrotron luminosity  increases with increasing electron injected power, and for  $L_{inj}=5\times 10^{38}$ erg s$^{-1}$ the system is synchrotron dominated. At the lower panel
we plot the analytical (broken line) and numerical (solid line) steady-state EED. At the upper panel we plot the analytical synchrotron (broken line) and numerical (solid line) synchrotron and SSC SED.  The four vertical lines mark four frequencies for which we study their variability in Figure \ref{fig:pulsetest} using the same line styles.
}
\label{fig:steadytest}
\end{figure}

A simple analytical test can be performed in the case of a source in which the energy losses are dominated by synchrotron radiation. In this case, adopting a steady power law electron injection at the inlet and assuming an electron residence time  $kL/c$ in the source, the source integrated EED will reach after time $t>>kL/c$ a steady-state.
 This  steady-state, source-integrated  electron distribution is a broken power law with an  electron index steepening by one  at $\gamma_b=(3 m_e c^2 )/(4\sigma_\tau U_B k L)$. This will produce a synchrotron spectrum with a spectral break of $1/2$ at an energy $\epsilon_b=B_\star \gamma_b^2$. 
For the synchrotron dominated configuration presented in Figure \ref{fig:steadytest}, the numerical result is in good agreement with the analytical both for the EED and the SED employing a $\delta$-function synchrotron emissivity. Note that the SSC luminosity is much lower than the synchrotron one. Note also that while the analytical synchrotron emission stops at 
$\epsilon_{min}=B_\star \gamma_{min}^2$, the numerical continues  to lower frequencies, due to the 
$f_\nu \propto \nu^{1/3}$ lower energy tail of the synchrotron emissivity. 

To compare the variability of our code with analytic expectations, we initiate injection of  a power low EED at $t=0$ at the inlet and  follow the evolution of the system toward a steady state.
We expect that the EED will reach a steady state  at or before a time $\sim k L/c=2 L/c $, equal to the time it takes for an electron to transverse the length of the pipe and escape, or equivalently the time it takes to fill the pipe with electrons. 
The part of the EED with $\gamma<\gamma_b$ will reach a steady state
 at $\sim  k L/c $,   because the electrons responsible for this emission transverse the entire pipe without cooling appreciably. 
 At higher energies,  $\gamma>\gamma_b$,  the electron radiative cooling lifetime  
$t=(3m_e c)/(4\sigma_\tau U_B\gamma)$ is shorter than $t=k L/c$. Electrons, therefore, of progressively higher energy will be confined  closer to the inlet, resulting to an energy-dependent size pipe.  This, together with the orientation of the observer, determines the  time it takes for the emission at a given energy to reach the steady-state. 

For an  observation angle $\theta=\pi/2$, there will be no position-dependent  delays, given that the light
path from all parts of the pipe to the observer are equal. Note that if the source is moving relativistically with bulk Lorentz factor $\Gamma$,  $\theta=\pi/2$ transforms to $\theta=1/\Gamma$ in the observers' frame.
At
 $\epsilon_s<\epsilon_b$ we expect a practically achromatic increase of the luminosity, reaching a steady state at $\sim  k L/c $  because the electrons responsible for this emission have $\gamma<\gamma_b$.
 At  higher energies $\epsilon_s>\epsilon_b$ the emission comes from 
electrons with energy $\gamma> \gamma_b$,  and the time to reach steady-state is  
$t=(3m_e c)/(4\sigma_\tau U_B\gamma)=(3m_e cB_\star^{1/2})/(4\sigma_\tau U_B \epsilon_s^{1/2})$,
where we have used $\epsilon_s=B_\star \gamma^2$.

\begin{figure}
\epsscale{0.5}
\plotone{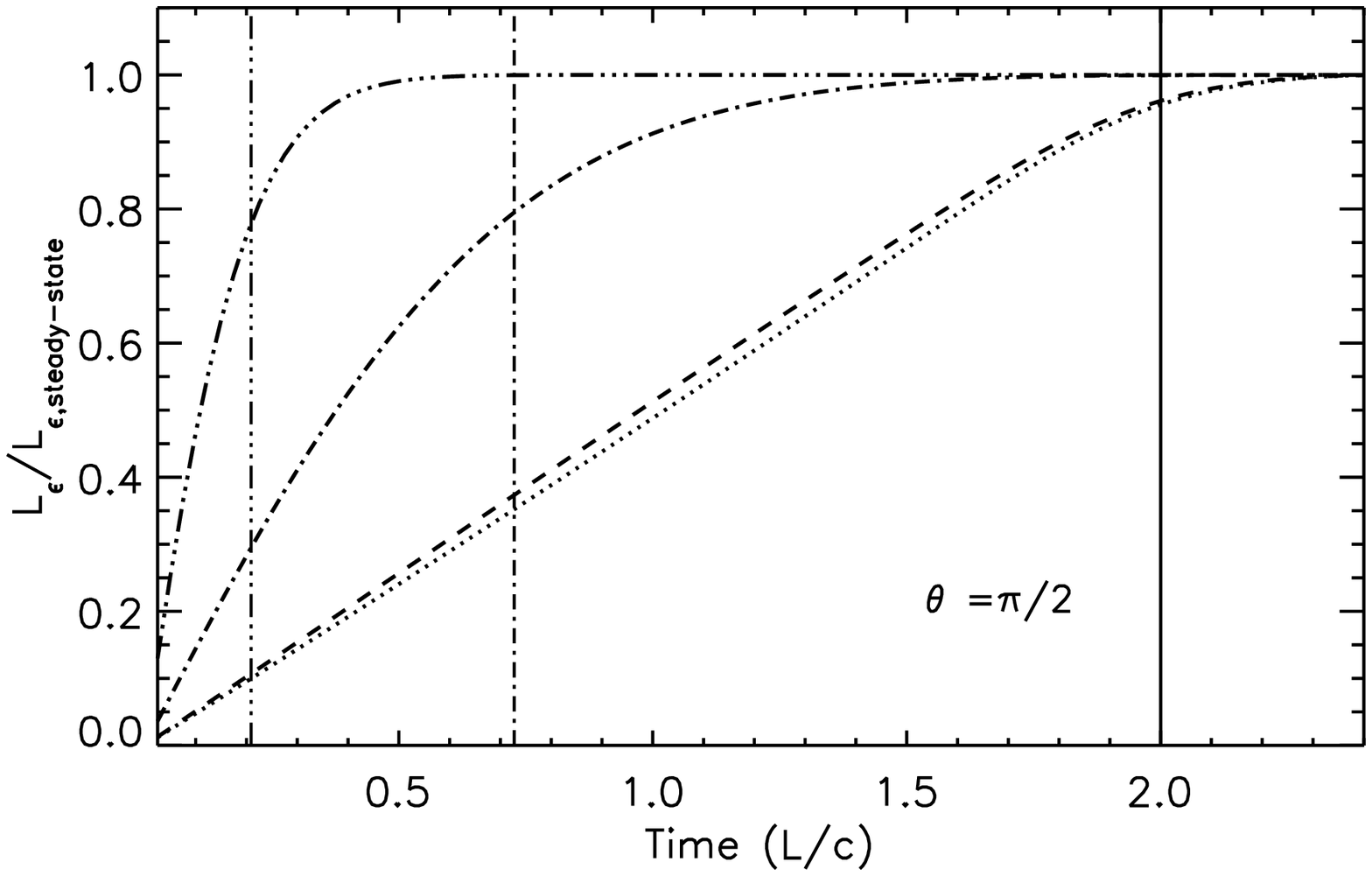}
\plotone{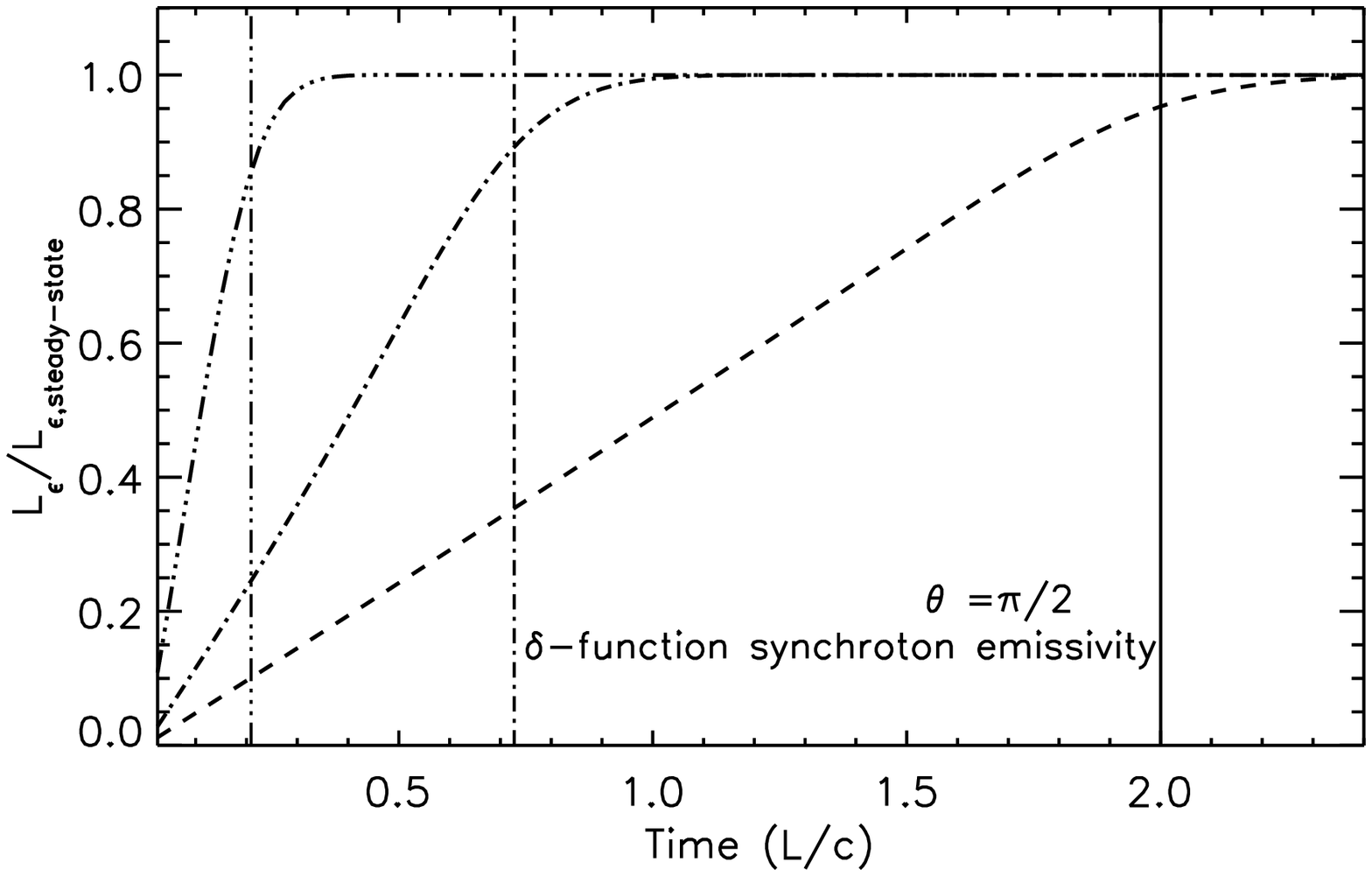}
\plotone{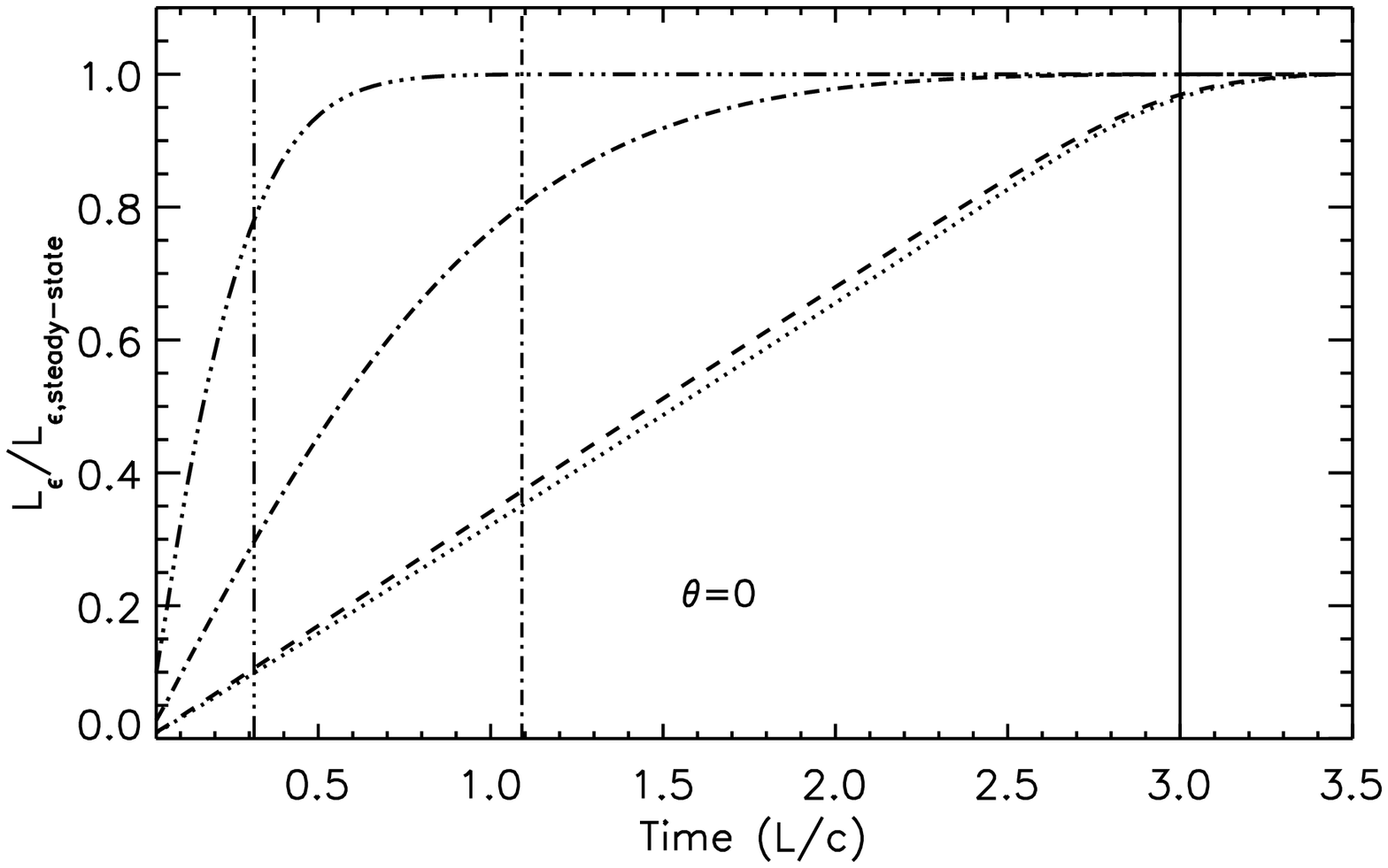}
\caption{The light curves of the four frequencies shown by the vertical lines in Figure \ref{fig:steadytest},
using the same line styles, for two different pipe orientations. The solid vertical lines correspond to the time $kL/c$ required  for the low energy ($\epsilon < \epsilon_b$) light curves to reach steady state,
while the dot-dash and triple dot-dash  vertical lines  correspond to the analytical estimate for the time it 
takes for the high energy ($\epsilon  >\epsilon_b$) light curves to reach steady state. The pipe orientation is $\theta=\pi/2$ for the upper and middle panels, and $\theta=0$ for the bottom panel.
The middle panel uses a $\delta$-function approximation for the synchrotron emissivity. }
\label{fig:pulsetest}
\end{figure}

 To verify that  the model variability
agrees with our analytical predictions, we select four synchrotron energies, marked through the four vertical lines in Figure \ref{fig:steadytest}. The lowest energy (dotted vertical line) comes entirely  from the $\nu^{1/3}$ tail of the synchrotron emissivity, and it is heavily dominated by the lowest  energy electrons with $\gamma=\gamma_{min}<<\gamma_b$ that have no time to cool before they escape. 
The second lower  energy (broken vertical line in Figure \ref{fig:steadytest})  at 
$\epsilon_{min}<\epsilon < \epsilon_{b}$ is predominately due to electrons with   $\gamma_{min}<\gamma < \gamma_{b}$ that also escape before they cool appreciably.
 In the upper panel of Figure \ref{fig:pulsetest} we plot the model light curves of these two low synchrotron energies, using the same line styles. As expected, the two light curves are almost indistinguishable, both reaching a steady state at $t\sim 2 L/c$ (marked by a solid vertical line in the upper panel of Figure \ref{fig:pulsetest}).

The light curves
of  two  more energies, this time  with  $\epsilon_s>\epsilon_b$,  are  marked by the dot-dash and triple dot-dash lines in Figure \ref{fig:steadytest} and 
are plotted in the upper panel of Figure \ref{fig:pulsetest}  using the same line styles. The vertical lines with the same line styles in Figure \ref{fig:pulsetest} indicate the time at which the corresponding light curves are expected to reach a steady state. 
As can be seen, at these times the light curves are at $\sim 80 \%$ of their steady-state level. 
This is mainly  because  electrons continue to radiate at a given energy $\epsilon_s$ even when their Lorentz factor  drops below $(\epsilon_s/B_\star)^{1/2}$, as  the exponential decay of the emissivity
 indicates (equation   \ref {eq:melrose}; e.g.   the synchrotron emissivity of an electron with Lorentz factor $\gamma= (\epsilon_s/B_\star)^{1/2}$ at time $t_0$ drops  $\propto \exp[ -(t/t_0)^2]$ from its peak emissivity,  requiring $t=2t_0$ to drop by $98\%$). To verifly this, we plot in the middle panel of Figure \ref{fig:pulsetest}  the same light curves, using the $\delta$-function approximation for the synchrotron emissivity. As can be seen, the two high energy light curves approach the steady-state level significantly closer to the analytically expected time.

To evaluate   if the light travel effects are properly taken into account, we rotate the pipe  in such a way that the inlet is closer to the observer ($\theta=0$), as is  the case for a propagating shock, that is  observed  `jet on'. In this case, we anticipate  the low energy  ($\epsilon < \epsilon_b$) variability that requires to fill the entire pipe with plasma, will reach a steady state after an additional time $L/c$, because this is the additional
length that variations from the end of the pipe have to travel to reach the observer. The time it takes, therefore, for the lower energy emission steady-state to be reached is $(k+1)L/c$. 
For higher energy variability ($\epsilon > \epsilon_b$) that reaches a steady-state before time  $t<kL/c$, the additional light travel time required is $(tc/k)/c=t/k $, where $tc/k$ is the maximum distance from the inlet of the pipe that the  contributes to the  energy under consideration. The time it takes, therefore, for the steady-state to be reached is $t(1+1/k$). We demonstrate these considerations in the lower panel of Figure \ref{fig:pulsetest}.
As can be seen the time it takes for the low energies ($\epsilon < \epsilon_b$) to reach steady state is now $(k+1)L/c=3L/c$, and the time it takes for the high energy light curves  to reach a steady state
increases by a factor $\sim 1/2$ as can be seen by comparing the lower and upper panels of Figure
\ref{fig:pulsetest} (e.g.  the dot-dash light curve reaches $95\%$ of the steady state after $t\approx 1.1\, L/c$ for $\theta=\pi/2$ and after $t\approx 1.65\, L/c$ for $\theta=0$).

  An analytical result regarding the relation of the synchrotron to the SSC emission,
 that we can test our model against,
   addresses  the relative amplitude of synchrotron and  SSC emission.
  For states that are synchrotron dominated, an increase of the electron injection normalization results to a linear increase of the synchrotron emissivity because the synchrotron emissivity is proportional to the number
  of available electrons, and to a quadratic increase of the SSC luminosity,  because the SSC luminosity
  is proportional to the product of  number of electrons times the synchrotron photon energy density, 
  which  scales with the number of electrons (e.g. Ghisellini, Maraschi,  \& Dondi 1996). This result holds as long as the   increased injection lasts long enough to occupy the entire volume of the source
  (in our case $t_{var}>kL/c$), bringing the source to a new steady state.
  
\begin{figure}
\epsscale{0.5}
\plotone{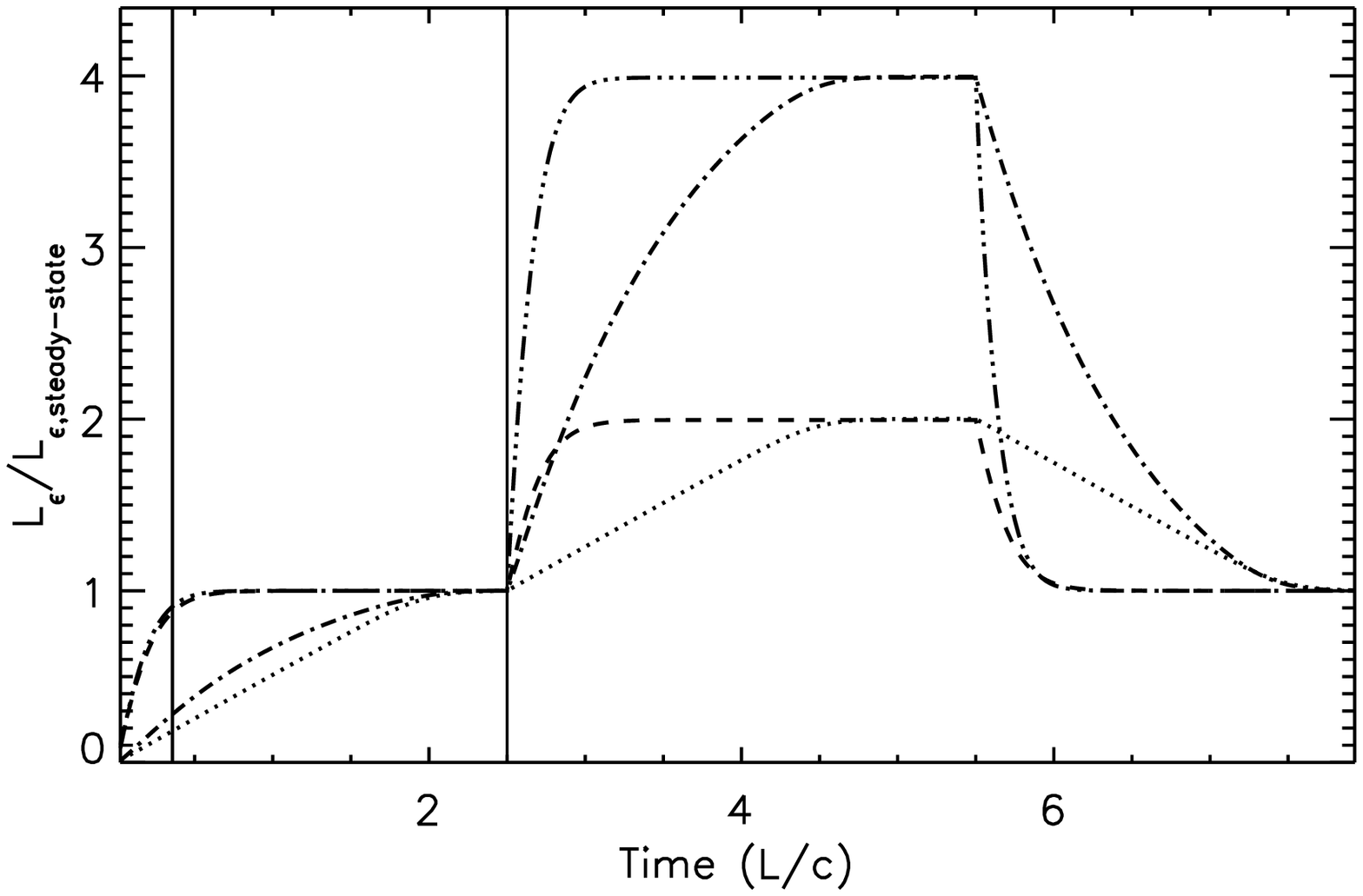}
\plotone{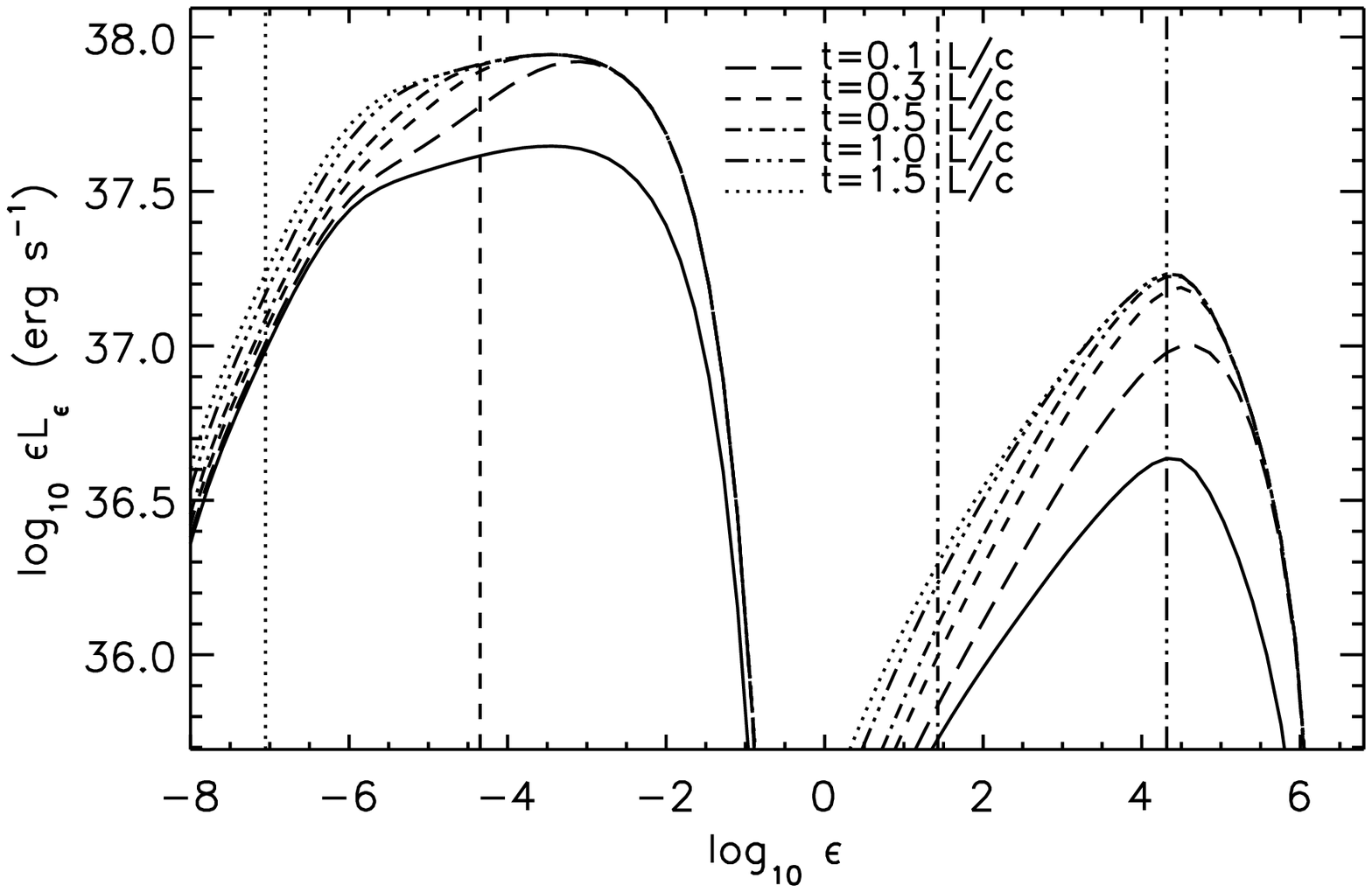}
\plotone{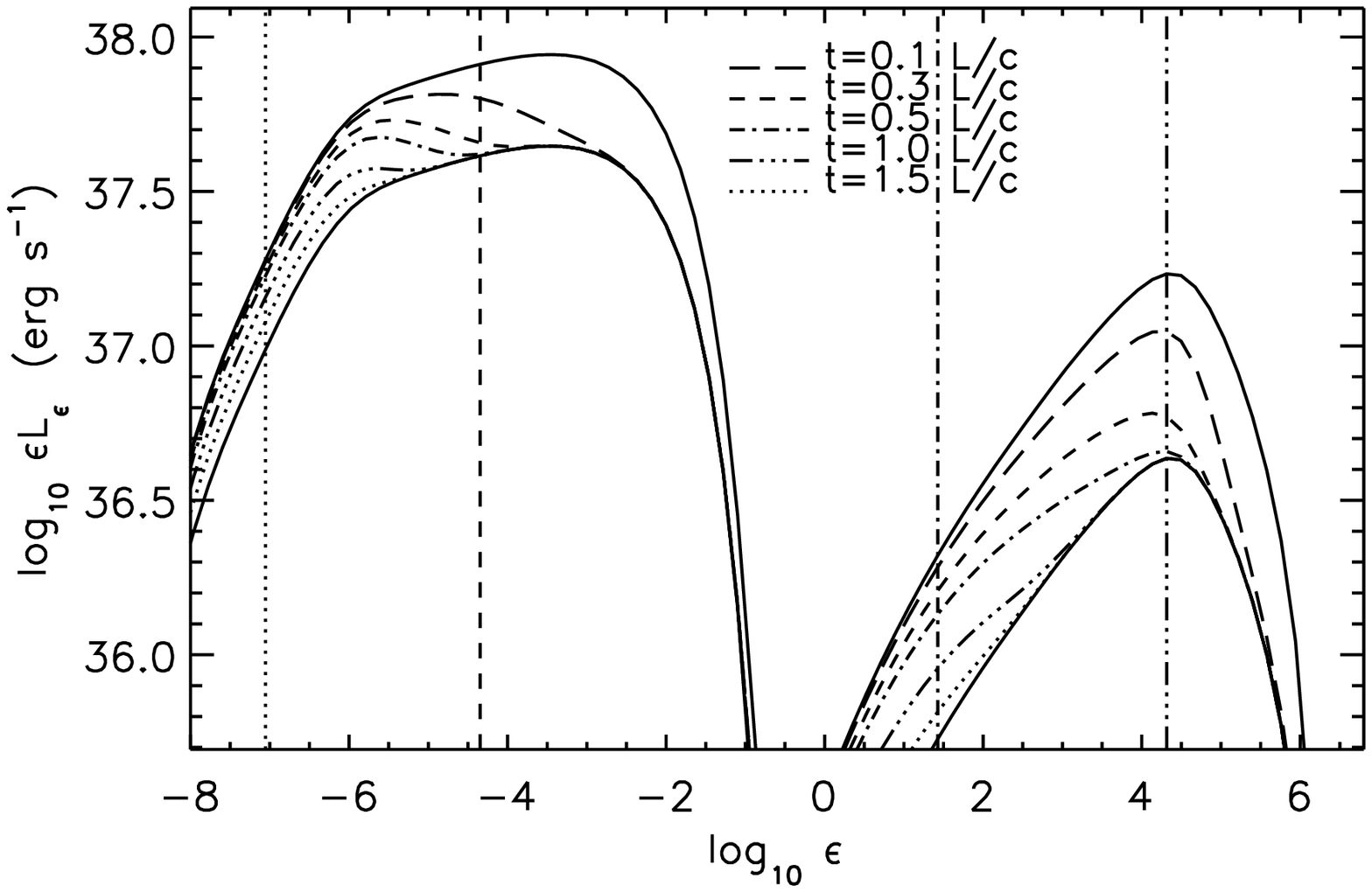}
\caption{A quadratic variation: a doubling of the injected electron power for $t=3L/c$,  a time longer than $2L/c$, the time it takes electrons to transverse the pipe.  The initial configuration is the same as that of Figure \ref{fig:steadytest}. The middle panel shows snapshots of the SED for a range of times elapsed from the beginning of the additional injection and the bottom  panel shows snapshots of the SED evolution after the additonal injection has been switched off.  The upper panel shows the light curves for the four energies depicted by vertical lines at the lower two panels.    }
\label{fig:quad}
\end{figure}

  To verify that our code reproduces this behavior we start from the configuration of Figure \ref{fig:steadytest}, which we observe from an angle $\theta=\pi/2$  and, after we let the system reach its steady state, we increase the injected electron luminosity by a factor of $2$ for
$t=3L/c$, longer than $2L/c$, the time to reach steady state. The middle panel of Figure \ref{fig:quad} shows the evolution of the SED as the flare grows, with the times denoting time since the increased injection started, while the 
bottom panel shows the evolution of the flare as the flare dies out, starting from the time the additional
injection is switched off.  The characteristic quicker response of the high energy electrons is apparent.
The two solid curves represent  the low and high steady-states.
At the upper panel we plot the light curves of the four frequencies marked with vertical lines at the two bottom panels. They are selected to roughly correspond to optical, X-ray, GeV and TeV energies,
assuming that beaming  will increase their observed values by a  Doppler factor $\delta\sim 20-40 $.
Note that the synchrotron emission doubles when it reaches its steady-state, while the SSC quadruples,
in agreement with the analytical result.

\subsection{Case studies \label{subsection:flares}}
\begin{figure}
\epsscale{0.5}
%\plotone{olopulse.eps}
\plotone{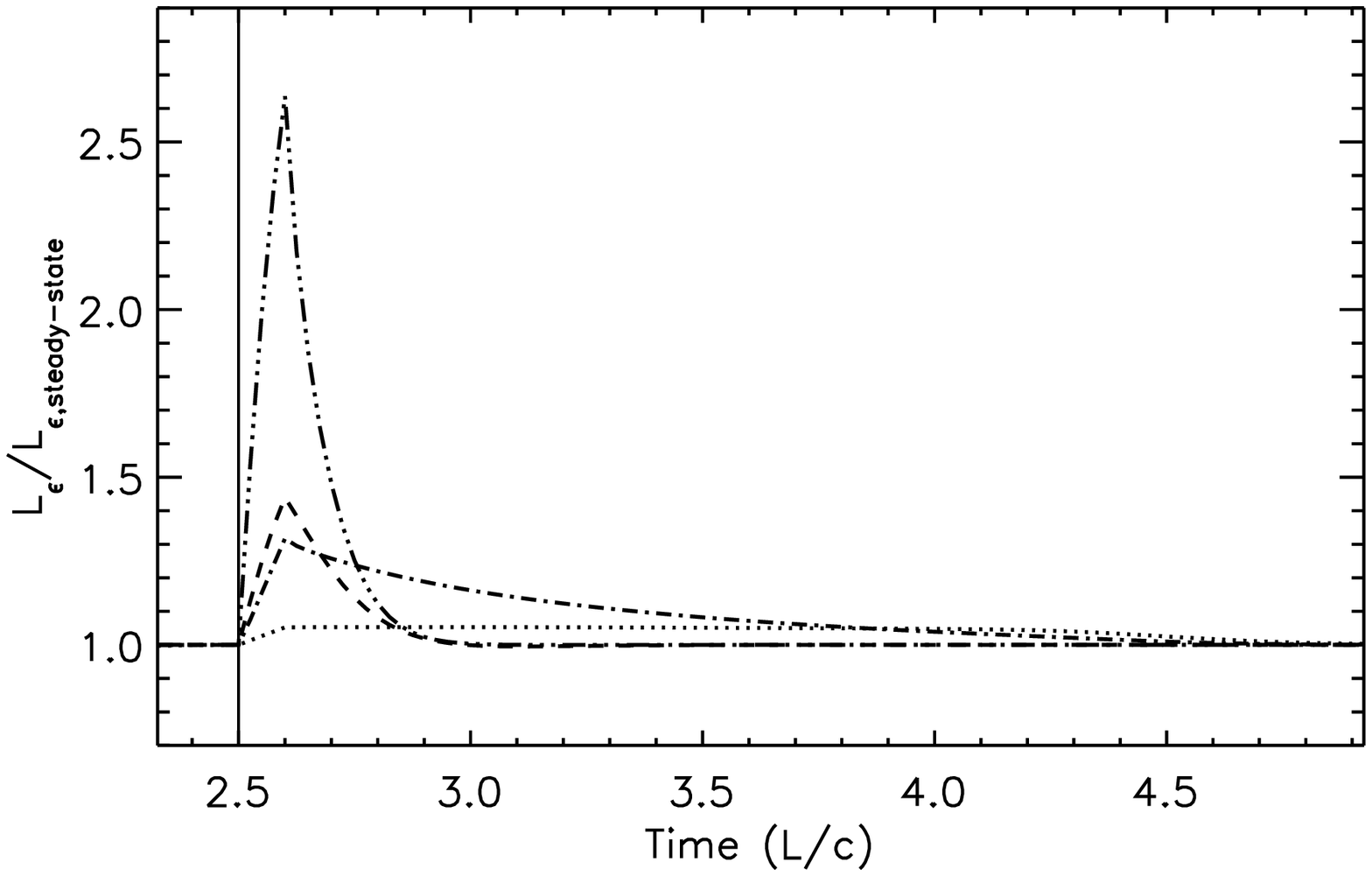}
%\plotone{sed_evol_olo.eps}
\plotone{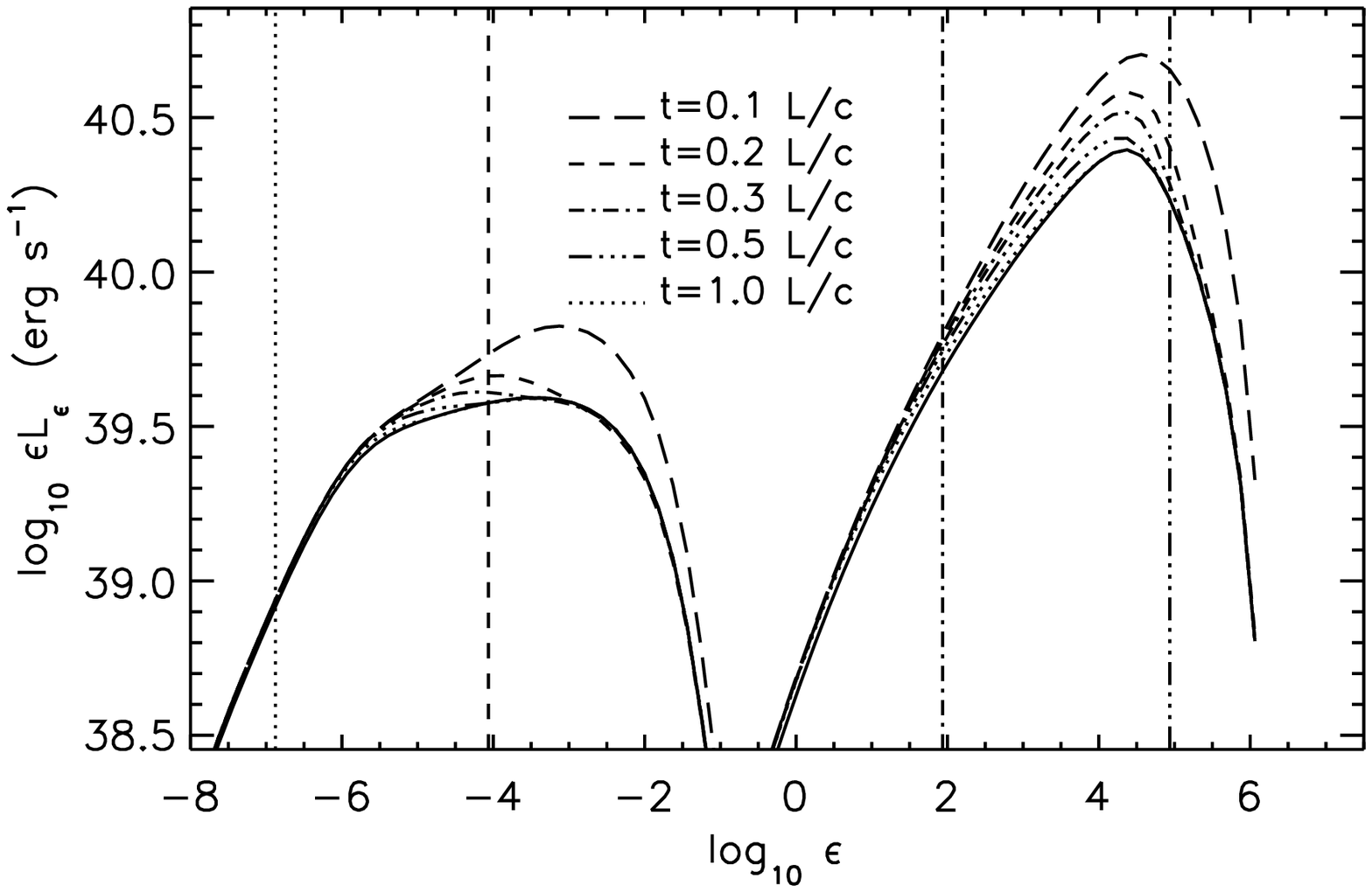}
\caption{Here we use the same steady-state configuration as in Figure \ref{fig:steadytest}, except for higher injected EED luminosity, $L_{inj}=5\times 10^{40}$ erg s$^{-1}$. After the system reaches the steady-state, we increase the injected luminosity by a factor of 2 for $0.1 L/c$. The bottom panel shows the evolution of the SED, while the upper panel the light curves of the flare for the four energies marked by the vertical lines at the bottom panel. }
\label{fig:olopulse}
\end{figure}

We present here three variability case studies,  for which we use as a starting point  the configuration  described in Figure  \ref{fig:steadytest}, increasing the electron
luminosity to $L_{inj}=5 \times 10^{40}$ erg s$^{-1}$ to produce a steady-state SED (solid line  in the lower panel of Figure \ref{fig:olopulse}) that  for a beaming $\delta\sim 20-40$  resembles those produced by flaring TeV blazars. 
We first study the case of an increased electron injection that  lasts a short fraction of the light crossing time.  
After the system reaches its steady state, we increase the injected electron luminosity $L_{inj}$ by a factor of $2$ for $0.1 \,L/c$.
At the bottom panel of Figure \ref{fig:olopulse}, we plot the SED evolution, while at the upper panel we plot the light curves that correspond to the four energies denoted by vertical lines at the bottom panel.
We note that the flare peaks with no time-delays at all energies, and then decays with the higher energies of each component decaying first. We also note that both the synchrotron and SSC  flare
have a higher amplitude at higher energies (and therefore the spectrum hardens as the flux increases).
 Also,  because additional electrons are injected at all
energies, the entire SED responds to the increased injection. The maximum fractional increase of the emission increases with frequency for both the synchrotron and the SSC components. This is because the higher the energy of the electrons required to produce a given synchrotron or SSC emission, the shorter their lifetime in the source; an additional injection, therefore,   for a fixed time ($0.1 L/c$ in our case) will increase by a higher factor the number of higher energy electrons. 

\begin{figure}
\epsscale{0.5}
\plotone{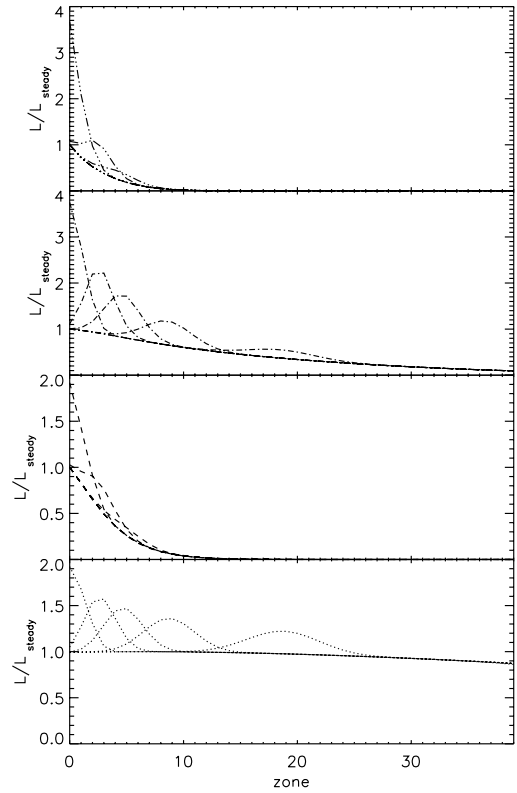}
\caption{The luminosity profile along the pipe for the four energies whose light curves are plotted in Figure \ref{fig:olopulse}, for the same pipe configuration. In each case the lower curve represents the luminosity profile, normalized to the steady-state  luminosity of the first zone. The profiles at times
$t=0.1,\, 0.2,\, 0.3,\, 0.5, \, 1.0$  light crossing times are also plotted and can be seen moving away from the inlet as the additional injection propagates.}
\label{fig:zones}
\end{figure}

To show how the variability
event would be seen if we had the ability to resolve the pipe, we plot in the four panels of Figure \ref{fig:zones} the luminosity profile along the pipe for the four energies we study. In all cases we normalize
the luminosity to the steady-state luminosity of the first zone. The lower curve in all cases depicts the steady-state luminosity profile. As expected, the high energy synchrotron and SSC emission are confined close to the inlet, while the low  energy emission of both components extends throughout the source. The decline of the steady-state low energy SSC emission along the pipe  is due  to the fact that
this emission is a convolution of a range of electron energies, 
and  the higher energy electrons are gradually becoming unavailable away from the inlet.
On top of the steady-states, we plot the snapshot  luminosity profiles at the times depicted in the lower panel of Figure \ref{fig:olopulse}. It can be seen how the pulse is propagating away from the inlet, gradually disappearing due to cooling at the  high synchrotron and SSC energies. 
At the  low synchrotron and SSC energies it can clearly be seen how the pulse is spreading out
due to the escape from zone to zone (the $n(\gamma,t)/t_{esc}$ term of the kinetic equation).
Note also that while the amplitude of the low energy synchrotron variation is substantial (starting with an increase by a factor of 2 at the inlet), because the pulse lasts for only $0.1 L/c$, while low energy synchrotron emission is produced by electrons accumulating for $2 L/c$, the increase of the total low energy synchrotron emission is small (about $5 \%$) as expected, and as can be seen in the upper panel of Figure  \ref{fig:olopulse}.

\begin{figure}
\epsscale{0.5}
\plotone{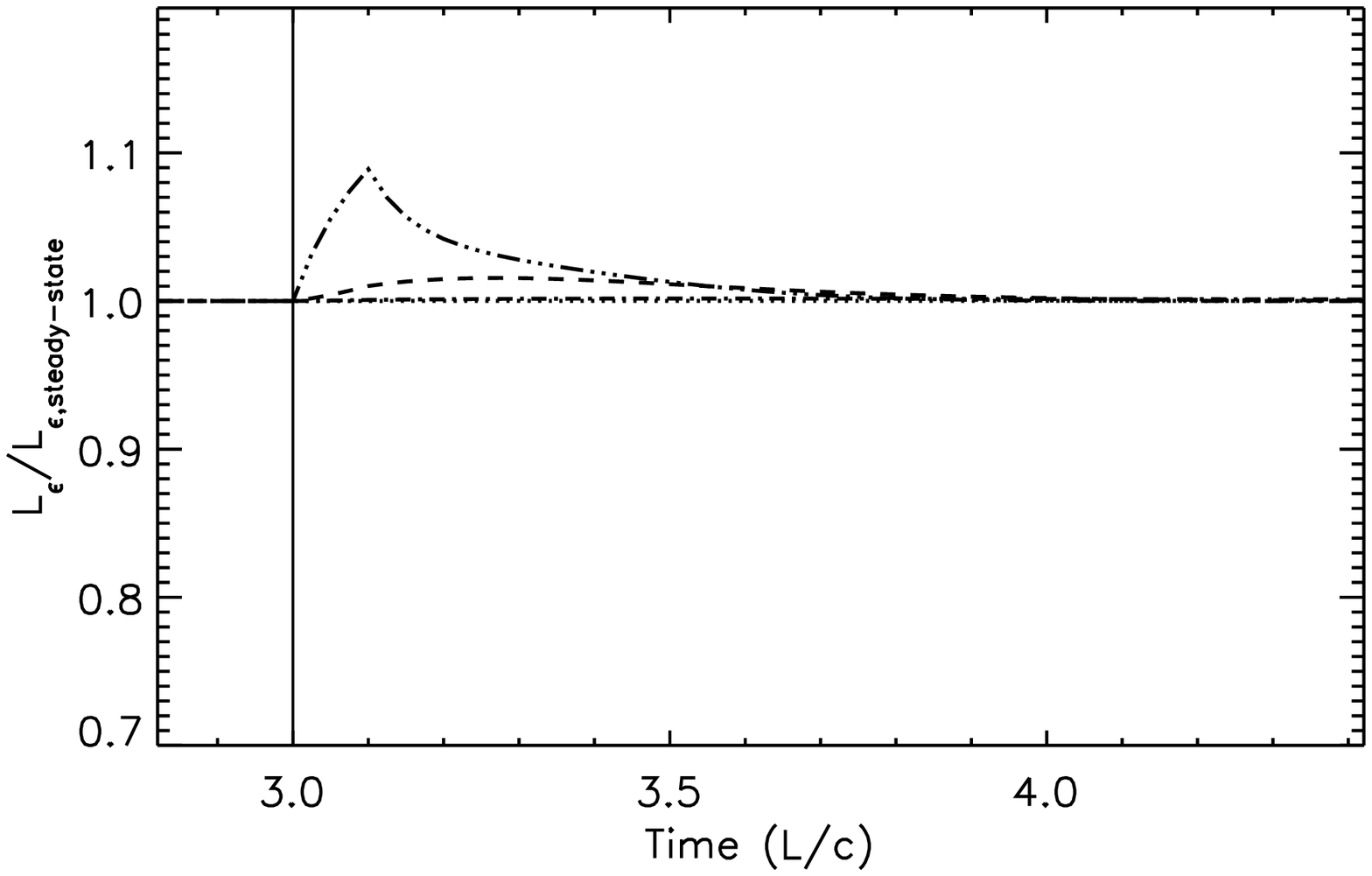}
\plotone{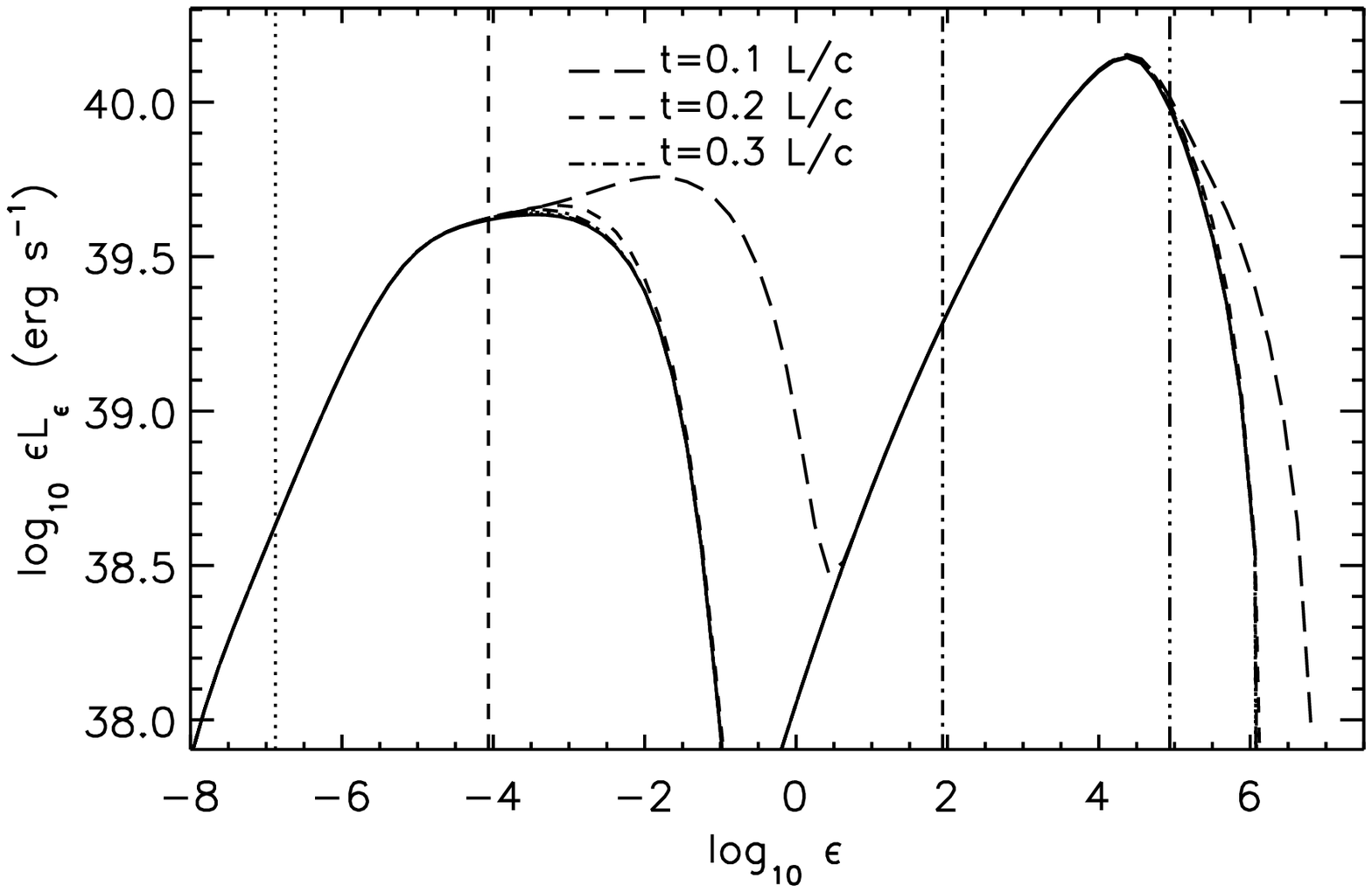}
\caption{Starting from the same steady state as  in Figure \ref{fig:steadytest}, $\gamma_{max}$ is increased by a factor of 5 for $0.1 L/c$.}
\label{fig:panopulse}
\end{figure}

Another  possible variation in the injected electron distribution is an increase in the maximum electron Lorentz factor of the EED, something that can result from  a temporary increase  in the electron acceleration rate. In this case the normalization of the EED remains constant and the increase in the injected luminosity depends on the electron index and the new value of $\gamma_{max}$.  Using  the same configuration as above,
we increase the value of $\gamma_{max}$ by a factor of 5 for the same short time $0.1 L/c$ (for an electron index of $p=1.8$ and for $\gamma_{min}=10^3$ this corresponds to an increase of $L_{inj}$ by $\sim 50\%$) and we follow the evolution of the SED. As can be seen in Figure \ref{fig:panopulse}, the event is mostly manifested at the high energy tails of both the synchrotron and SSC components and dies quickly, because the high energy electrons injected have very short lifetimes.
No significant  variations are seen away from the high energy tails of the synchrotron and SSC components, a notable difference from the previous case in which the normalization of all electrons was increased.

Finally, we study a case in which after the steady state is reached,  an additional population of relatively  low energy electrons is injected for a short time. This may be a plausible situation if one considers that
a  pre-acceleration mechanism is required to  accelerate electrons  up to Lorentz factors $\gamma\sim  \Gamma m_p/m_e$, where $\Gamma$ is the typical bulk Lorentz factor of the flow,  to provide the electrons that can be picked up by the Fermi acceleration mechanism for acceleration to much higher Lorentz factors (e.g. Sikora et al. 2002). Variations in this pre-acceleration mechanism that are not propagated to  Fermi acceleration may account for the variations in the low energy tail of the electron distribution.
We simulate this scenario by injecting an additional low energy EED with the same luminosity as the steady injection, but with $\gamma_{max}=10^4$. 

\begin{figure}
\epsscale{0.7}
%\plotone{finalorfan1varpaper.eps}
\plotone{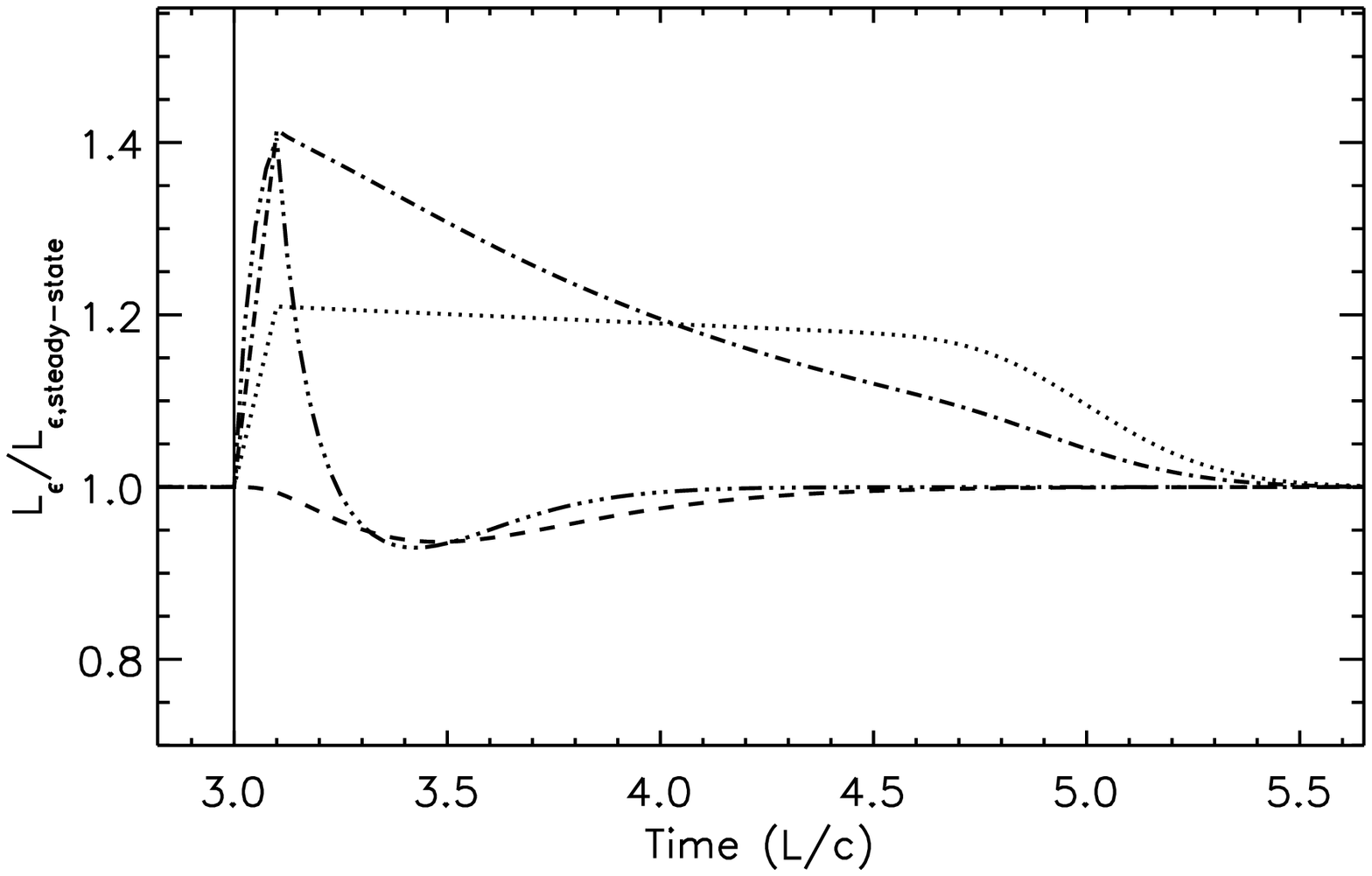}
\plotone{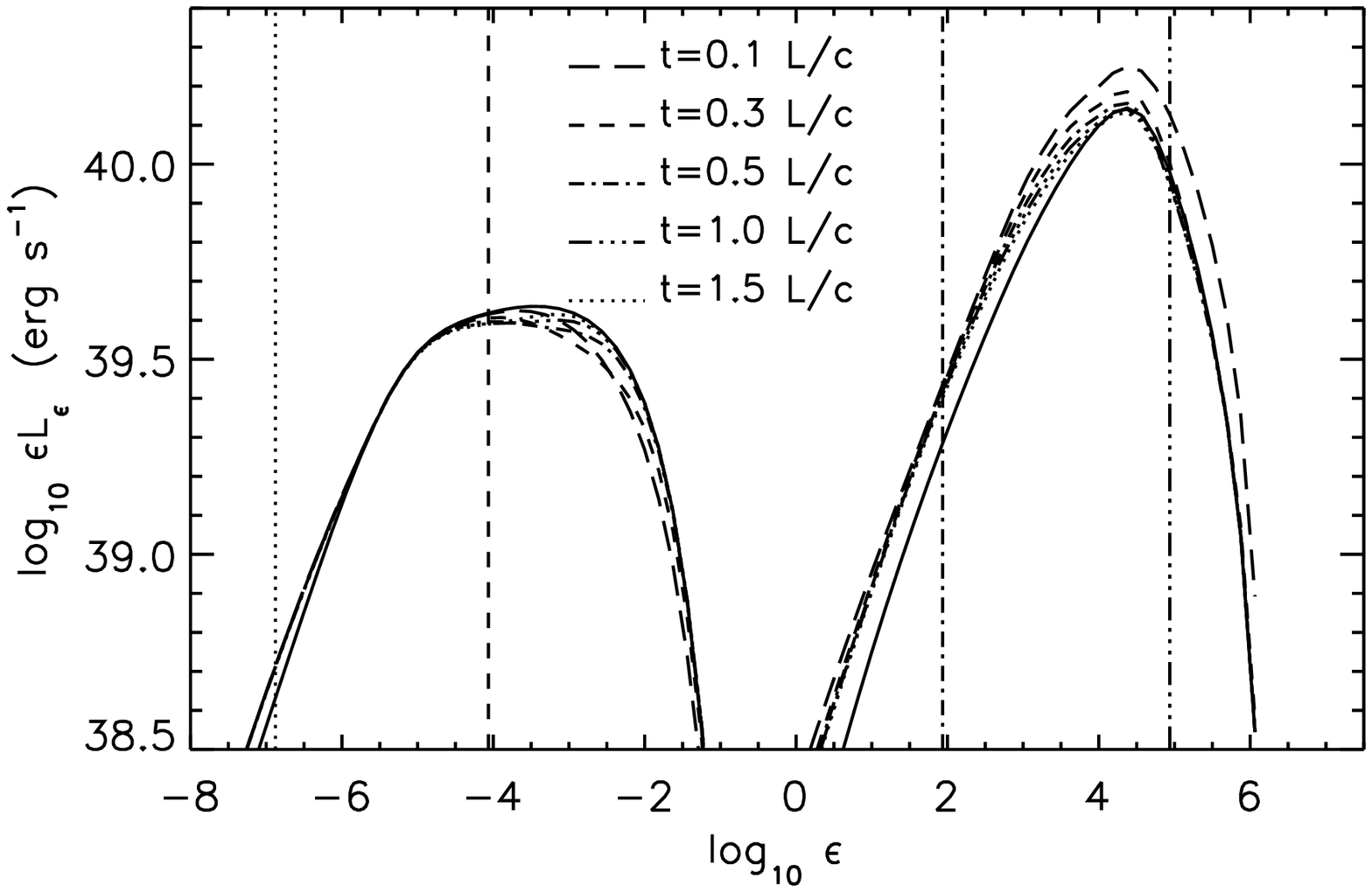}
%\plotone{f2.eps}
\caption{Starting from the same steady state as  in Figure \ref{fig:steadytest}, we inject for $0.1 L/c$ an additional low energy EED with the same luminosity as the steady injection, but with $\gamma_{max}=10^4$.}
\label{fig:flare1}
\end{figure}

As can be seen in  Figure \ref{fig:flare1}, as soon as the injection starts, the additionally injected  low energy electrons produce  low energy synchrotron emission  (dotted line in Figure \ref{fig:flare1}) that remains at a plateau during the time it takes for the extra injection to transverse the length  of the pipe, because these low energy electrons do not cool strongly.
In the beginning, these low energy synchrotron photons are  produced close to  the TeV energy  electrons of the steady flow (due to radiative losses these high energy electrons are found only close to the injection inlet)  and are upscattered by them to  TeV energies, producing additional TeV emission which is manifested as the rising part of the TeV flare (triple dot-dash line).
As the event evolves,  the additional population of  
low energy electrons propagates downstream and, due to the increasing
distance between the additionally  produced synchrotron photons and the pipe inlet where the TeV electrons are found, the TeV flare quickly subsides.
The behavior of the X-ray emission  (broken line) is very interesting: the extra low energy synchrotron photons produced increase the photon density experienced by the synchrotron X-ray
emitting electrons and this  causes additional cooling, slightly reducing   the X-ray synchrotron emission.  
As the production of these additional low energy synchrotron photons is displaced downstream, their
effect at the neighborhood of the inlet, where the synchrotron emitting electrons are found, decreases
and the X-ray flux returns to its steady state.
The behavior of the lower energy Gamma-ray emission does not show the sharp decline of the high energy Gamma rays. Instead, their light curve shows a gradual decline. This flare is mostly produced by the additional electrons upscattering  a broad energy range of seed photons that has a gradually decreasing level with distance from the inlet energy. 
The model behavior of a TeV flare not accompanied by an X-ray flare is reminiscent  of the so-called 
 orphan TeV flares \citep{krawczynski04,blazejowski05}.  
These are rare flaring states of blazars characterized by an increase in the TeV luminosity  that is not accompanied by a similar increase in X-ray energies.

\section{Discussion and future work\label{section:discussion}}
We presented  a multizone code
that for the first time takes into account the  non-local, time-retarded nature of SSC losses. 
 This  code is currently the only multizone model that incorporates the non-local, time-delayed  SSC  losses, and as such is uniquely suitable for modeling the results of  multiwavelength campaigns  at radio, optical, X-ray and $\gamma$-ray energies, with the additional constraints in the critical and unexplored  for TeV blazars GeV  GLAST regime.

As we argued, the results of one zone codes for the critical high energy regime of both the synchrotron and SSC components are problematic, and should not be used to infer the physical conditions in the source through variability modeling.  We described our multizone code,  tested it successfully against known analytical results, and presented a small number of variability case studies. 
The case studies we presented, although based on the same underlying steady-state configuration, exhibited very different variability patterns. This means that detailed modeling of 
 broadband SEDs and simultaneous multiwavelength variability can be used to infer what is actually the cause of a given observed variability pattern, providing reliable constraints on the particle acceleration taking place.  Orphan flares can be reproduced assuming an increase of the injection of the low energy electrons, but not assuming  the injection of a very high energy electron population, as we also showed
 analytically.   
 
 The fact that this plausible variation  cannot produce orphan flares significantly   narrows the   parameter space for events that can produce such events, possibly in agreement with their observed scarcity.  

The  code we described  can  run  with a typical workstation in a reasonable time of at most a few minutes at a resolution of $\sim 10$ bins per decade of observing frequency, $\sim 10$ bins per decade
of electron energy, and $\sim 50$ zones.
 To achieve this we  employed a pipe geometry, and  adopted an energy conserving $\delta$-function approximation for the  SSC  emissivity, as well as a step function approach to take into account the change from the Thomson to Klein-Nishina IC scattering cross-sections.
Adopting  these approximations is problematic  for situations where IC scattering of narrow photon distributions (e.g. line emission from the broad line region or even a blackbody spectrum characterized by a typical photon energy $\epsilon_0$)  
is important. In this case the adoption of the step function cross section description would create a strong artificial feature on the EED localized at the transition
from the Thomson to the Klein-Nishina regimes at $\gamma\propto 1/\epsilon_0$, which would then 
propagate to the emitted spectra through the $\delta$-function IC emissivity. 
For SSC systems, however, where  the seed photons are spread over many decades in energy,
the resulting spectra are good approximations of those   produced using the full expressions for the synchrotron and SSC emissivities as well as the full Klein-Nishina cross section.

Including the above considerations, as well as the processes of synchrotron opacity and pair production through $\gamma$-ray absorption  within the source, would  increase the execution time up to levels marginally comfortable for the typical workstation. Most probably, such an extension of the code would require parallelization.  A more desirable upgrade of the code would drop the assumption of no lateral gradients in  the plasma characteristics by  switching to a two-dimensional geometry, in which 
the electron distribution and the SSC photon energy density are allowed to change laterally to the flow direction. Such considerations may be relevant to the recently observed $\sim   0.75 $ days  delay between the IR and the X-ray variability in 3C 273 \citep{mchardy07}.    These authors argued that
the delay may be attributed to the time it takes for the SSC photon energy density to built up as the SSC photons are transversing   the cross section of the flow. This upgrade will scale the computation time roughly by $N^2$, increasing it  from $\sim$  few minutes
to  $\sim$  several hours. We note here that our formalism can be extended to  treat velocity profiles
in term of the decelerating flow \citep{georganopoulos03} or the spine sheath model \citep{ghisellini05}
that have developed to address the lack of superluminal motions in TeV blazars (e.g. Piner \& Edwards 2004).

Another upgrade that can be incorporated in the existing code, this time with a minimal computational overhead, is that of a zone for particle acceleration, following the formalism of \cite{kirk98}. In this case, in the first zone of the model, low energy  electrons will be injected and allowed to accelerate
while suffering radiative losses due to synchrotron and non-local SSC. These particles will subsequently
escape into the pipe and flow downstream. This configuration will require a different numerical scheme for the acceleration zone, since there most particles are advected upward in energy space, but there is a possibility, in a time-dependent scenario, of the highest energy particles being advected downward, while the rest of the electrons are still advected upwards. The benefit of including particle acceleration in the code is that it will allow us to study cases of hard lags/ counterclockwise loops in the X-ray hardness - X-ray flux diagrams thought to result when acceleration and loss timescales are comparable. Such a  code could  be used to model the observed curved X-ray spectra of high peak frequency blazars
in the framework of episodic particle acceleration \citep{perlman05}.

\acknowledgements

We are grateful to the referee for a critical review of the first version of the manuscript that made us aware of some important limitations of the first version of our code.
Part of this work was done in the context of the senior thesis of Philip Graff at UMBC.
%The  code can be found at  {\sl http://jca.umbc.edu/$\sim$markos/cs}.
The authors acknowledge support from NASA LTSA grants \# NAG5-9997 and NNG05GD63G at UMBC, as well as \#NNX07AM17G at FIT, and from the  Chandra theory grant TM6-7009A  at UMBC.

%\pagebreak

\clearpage

\begin{thebibliography}{99}



\bibitem[Aharonian (2000)]{aharonian00} Aharonian, F. A. 2000, New Astronomy,  5,  377

\bibitem[Aharonian et al. (2005)]{aharonian05} Aharonian, F. A., et al. 2005, \aap, 442, 895 

\bibitem[Aharonian et al. (2007)]{aharonian07} Aharonian, F. A., et al. 2007, \apj,  664, L71

\bibitem[Begelman, Fabian, \& Rees (2008)]{begelman08} Begelman, M.  C., Fabian, A. C., \&  Rees, M. J. \mnras, 384, L19

\bibitem[Blandford (1978)]{blandford78} Blandford, R. D. 1978, in Pittsburgh Conference on BL Lac Objects, ed. A. N. Wolfe (Pittsburgh: Univ. Pittsburgh Press), 328

\bibitem[B{\l }a\.zejowski et al. (2000)] {blazejowski00} B{\l }a\.zejowski, M., Sikora, M., Moderski, R.,\&  Madejski, G. M. 2000, \apj, 545, 107

\bibitem[B{\l }a\.zejowski et al. (2005)] {blazejowski05}
 B{\l }a\.zejowski, M.  et. al. 2005, ApJ,  630,  130
 
 %\bibitem[Blundell et al. (2007)]{blundell07}  Blundell, K.  M.,  Fabian, A. C., Crawford, C. S.,  Erlund, M. C.,\&  Celotti, A. 2007, \apj, 644, L13

\bibitem[B\"ottcher (2007)]{boettcher06} B\"ottcher, M. 2007, Astrophysics and Space Science, 309, 95

\bibitem[Crusius \& Schlickeiser (1986)]{crusius86} Crusius, A. \&  Schlickeiser, R. 1986, \aap, 164, L16

\bibitem[Chang \& Cooper (1970)]{chang70} Chang, J. S. \& Cooper, G. 1970, J. Comp. Phys., 6, 1
 
\bibitem[Chiaberge \& Ghisellini (1999)]{chiaberge99} Chiaberge, M.  \& Ghisellini, G. 1999, MNRAS  306,  551


\bibitem[D'Arcangelo et al. (2007)]{darcangelo07} D'Arcangelo, F. D. et al. 2007, \apj, 659, L107

%\bibitem[Fan, Y.Z. et al. (2008)]{fan08}	 Fan, Y.-Z., Piran, T., Narayan, R., \&  Wei, D.-M. 2008, \mnras, 384, 1483

 \bibitem[Fossati et al. (2000)]{fossati00} Fossati, G. et al. 2000, \apj, 541, 153

 \bibitem[Fossati et al. (2008)]{fossati08} Fossati, G. et al. 2008, \apj, 677, 906

\bibitem[Georganopoulos \& Marscher (1998a)]{georganopoulos98a} Georganopoulos, M. \& Marscher, A. P. 1998, \apj, 506, 621

\bibitem[Georganopoulos \& Marscher (1998b)]{georganopoulos98b} Georganopoulos, M. \& Marscher, A. P. 1998, \apj, 506, L11

\bibitem[Georganopoulos \& Kazanas (2003)]{georganopoulos03} Georganopoulos, M. \& Kazanas, D. 2003, \apj, 594, L27

\bibitem[Ghisellini, Maraschi,  \& Dondi, (1996)]{ghisellini96} Ghisellini, G.,  Maraschi, L., Dondi, L. 1996, A\&AS, 120, 503
 
\bibitem[Ghisellini, Tavecchio \&  Chiaberge (2005)]{ghisellini05} Ghisellini, G., Tavecchio, F. \& Chiaberge, M. 2005, \aap, 432, 401

\bibitem[G\'omez et al. (1994)]{gomez94} G\'omez, J. L.,  Alberdi, A., Marcaide, J. M.,  Marscher, A. P. \& Travis, J. P. 1994, \aap, 292, 33

\bibitem[Graff et al. (2007)]{graff07} 	 Graff, P. B., Georganopoulos, M., Perlman, E. S., Kazanas, D. 2007, The first GLAST symposium,  AIP Conference Proceedings, 921, 333

\bibitem[Harding \& LAi (2006)]{harding06} Harding, A. K. \& Lai, D. 2006, Rep. Prog. Phys., 69, 2631

\bibitem[Kardashev (1962)]{kardashev62} Kardashev, N. S.  1962, Soviet Astronomy,  6, 317 

\bibitem[Kataoka et al. (2000)]{kataoka00} Kataoka, J., Takahashi, T., Makino, F., Inoue, S., Madejski, G. M., Tashiro, M., Urry, C. M., \& Kubo, H. 2000, \apj, 528, 243

\bibitem[Katarzy\"{n}ski et al. (2006)]{katarzynski06} Katarzy\'nski, K., Ghisellini, G., Tavecchio, F.,  Gracia, J., \&  Maraschi, L. 2006, MNRAS, 368, L52

\bibitem[Kirk,  Rieger,  \&  Mastichiadis (1998)]{kirk98} Kirk, J. G., Rieger, F. M., \&  Mastichiadis, A.  1998, A\&A, 333, 452

%\bibitem[Kirk \& Tsang (2006) ] {kirk06} Kirk, J. G., \&  Tsang, O.  2006, \aap, 447, L13
 
\bibitem[Krawczynski,  Coppi, \&  Aharonian  (2002)]{krawczynski02} Krawczynski, H.,  Coppi, P. S., \&  Aharonian, F.  2002, \mnras,  336, 721
 
\bibitem[Krawczynski et al  (2004)]{krawczynski04}  Krawczynski, H.  et. al. 2004, \apj ,  601, 151 

%%\bibitem[Krennrich et al. (2002)]{krennrich02} Krennrich, F. et al. 2002, \apj, 575, L9

\bibitem[Maraschi, Ghisellini,  \& Celotti (1992)] {maraschi92} Maraschi, L., Ghisellini, G., \& Celotti, A. 1992, \apj, 397, L5

\bibitem[Maraschi et al. (1999)] {maraschi99} Maraschi, L. et al. 1999, \apj, 526, L81

\bibitem[Marscher et al. (2008)]{marscher08} Marscher, A. P.  et al. 2008, \nat, 452, 966

\bibitem[Mastichiadis \& Kirk (1997)] {mastichiadis97} Mastichiadis, A. \& Kirk, J. G. 1997, \aap, 320, 19

\bibitem[McHardy et al. (2007)]{mchardy07} McHardy, I.,  Lawson, A.,  Newsam, A. Marscher, A. P., Sokolov, A. S., Urry, C. M. Wehrle, A. E. 2007, \mnras,  375, 1521

\bibitem[Melrose (1980)]{melrose80} Melrose, D. Plasma AStrophysics Vol I, Gordon \& Breach, New York

\bibitem[Perlman et al. (2005)]{perlman05} Perlman, E. S. et al. 2005, \apj, 625, 727 

\bibitem[Piner \& Edwards (2004)]{piner04} Piner, B. G. \&  Edwards, P. G. 2004, \apj, 600, 115


\bibitem[Piner,  Pant,  Edwards (2008)]{piner08} Piner, B. G., Pant, N.,   \&  Edwards, P. G. 2008, \apj, 678, 64


\bibitem[Ravasio et al. (2004)]{ravasio04} Ravasio, M., Tagliaferri, G., Ghisellini, G., \&  Tavecchio, F. 2004, \aap, 424, 841

\bibitem[Sambruna et al. (2000)] {sambruna00} Sambruna, R. M. et al. 2000, \apj, 538, 127

\bibitem[Sikora, Begelman, \& Rees (1994)] {sikora94} Sikora, M., Begelman, M. C., \&  Rees, M. J. 1994, \apj,  421, 153


\bibitem[Sikora et al.  (2002)] {sikora02} Sikora, M.,  B{\l }a\.zejowski, M., Moderski, R., Madejski, G. M. 2002, \apj,  577, 78

\bibitem[Sokolov, Marscher, \& McHardy (2004)]{sokolov04} Sokolov, A., Marscher, A. P., \& McHardy, I. M.  2004, \apj, 613, 725

\bibitem[Sokolov \& Marscher (2005)]{sokolov05} Sokolov, A. \& Marscher, A. P.  2005, \apj, 629, 52

%\bibitem[Stawarz et al. (2007)]{stawarz07} Stawarz, L. Cheung, C. C.,  Harris, D. E.,  \& Ostrowski, M.
%2007, \apj, 622, 213

\bibitem[Takahashi et al. (1996)]{takahashi96} Takahashi, T.  et al. 1996, \apj, 470, L89

%\bibitem[ Tsang \& Kirk (2007)] {tsang07} Tsang, O.  \& Kirk, J. G.  2007, \aap, 463, 145


\bibitem[Zhang (2002)]{zhang02} Zhang, Y. H. 2002, \mnras, 337, 609

\end{thebibliography}
\end{document}